\documentclass[manuscript,screen]{acmart}
\AtBeginDocument{%
  }

\usepackage{wrapfig}
\usepackage{needspace}   
\usepackage{placeins}    
\usepackage{makecell}
\usepackage{hyperref}

\usepackage{array}
\usepackage{tabularx}   
\usepackage{booktabs}   
\usepackage[table]{xcolor} 
\usepackage{xspace}     
\usepackage{marvosym} 

\usepackage{enumitem}

\newcommand{\RQ}[1]{
    \texorpdfstring{\textbf{RQ\textsubscript{#1}}}{RQ#1}%
}

\newcolumntype{L}[1]{>{\raggedright\arraybackslash}p{#1}} 
\newcolumntype{Y}{>{\raggedright\arraybackslash}X}        

\newcommand{\opt}{\xspace\xspace\(\square\)\xspace} 

\definecolor{quotecolor}{HTML}{218380} 
\newcommand{\simplequote}[2]{\textit{\textcolor{quotecolor}{``#1''}$_{#2}$}}

\newcommand{\supplMaterial}{\url{https://github.com/AI-4-SE/Attitudes-and-Practices-Towards-Optimizing-Software-Energy-Consumption}}

\definecolor{rowgray}{gray}{0.95}

\usepackage{subcaption}
\usepackage{pdflscape}
\usepackage{ltablex}
\keepXColumns
\usepackage{rotating}
\usepackage{adjustbox}

\newcommand{\rowdesc}[1]{%
  \makebox[\linewidth][c]{%
    \smash{%
      \raisebox{-0.77\height}{%
        \rotatebox{90}{\parbox{2.9cm}{\centering\small\textit{#1}}}%
      }%
    }%
  }%
}

\usepackage{array}
\usepackage{graphicx}

\newcolumntype{C}[1]{>{\centering\arraybackslash}m{#1}}

\usepackage[commandnameprefix=ifneeded]{changes}

\begin{document}

\title{Developer Attitudes and Practices Towards Optimizing Software Energy Consumption}

\author{Max Weber}
\email{max.weber@informatik.uni-leipzig.de}
\orcid{0000-0003-3454-4121}
\affiliation{%
  \institution{Leipzig University}
  \city{Leipzig}
  \country{Germany}
}

\author{Alina Mailach}
\email{alina.mailach@informatik.uni-leipzig.de}
\orcid{0000-0001-6204-2095}
\affiliation{%
  \institution{ScaDS.AI Dresden/Leipzig, Leipzig University}
  \city{Leipzig}
  \country{Germany}
}

\author{Florian Sattler}
\email{sattlerf@cs.uni-saarland.de}
\orcid{0000-0003-2523-1158}
\affiliation{%
  \institution{Saarland Informatics Campus, Saarland University}
  \city{Saarbrücken}
  \country{Germany}
}

\author{Sven Apel}
\email{apel@cs.uni-saarland.de}
\orcid{0000-0003-3687-2233}
\affiliation{%
  \institution{Saarland Informatics Campus, Saarland University}
  \city{Saarbrücken}
  \country{Germany}
}

\author{Norbert Siegmund}
\email{norbert.siegmund@informatik.uni-leipzig.de}
\orcid{0000-0001-7741-7777}
\affiliation{%
  \institution{ScaDS.AI Dresden/Leipzig, Leipzig University}
  \city{Leipzig}
  \country{Germany}
}
\renewcommand{\shortauthors}{Weber et al.}



\begin{abstract}
\textbf{Context:} 
Software significantly influences the efficiency with which hardware resources are utilized, yet software energy consumption is seldom treated as a first-class concern in day-to-day development practice. 

\noindent
\textbf{Objective:} This study investigates professional developers’ attitudes, decision-making, and development practices related to software energy consumption, with particular emphasis on how energy considerations are recognized, assessed, and acted upon during software development.

\noindent
\textbf{Method:} To this end, we conduct an online survey with 134 software developers. Our study combines quantitative analyses with a qualitative open-card sorting of free-text responses to characterize perceptions, practices, and reasoning patterns around energy consumption.

\noindent
\textbf{Findings:} Energy consumption is explicitly considered in only a minority of projects. More commonly, developers influence energy use indirectly by optimizing proxy properties such as execution time and CPU utilization. 
Responses to scenario-based questions reveal systematic blind spots in this mental model, including cases in which performance improvements increase energy consumption or exhibit no correlation. We also identify organizational disincentives, limited tooling, and educational gaps as major barriers to adoption.

\noindent
\textbf{Implications:} (1) Institutionalize energy-aware approaches through visible flagship deployments that demonstrate value, (2) expand research and education on energy–-performance trade-offs, and (3) develop practical, developer-oriented measurement and feedback tools that lower adoption barriers.
\end{abstract}





\maketitle

\section{Introduction}
\label{intro}




Operating large-scale data centers~\cite{lei2021global,urgaonkar2010dynamic,beloglazov2012energy}, cloud infrastructures~\cite{procaccianti2013energy,beloglazov2012energy,arroba2014server}, and billions of connected devices~\cite{silva2017invasive,corbalan2018development,hao2013estimating} creates substantial power and cooling demand, driving high operating costs and environmental impact. 
Although energy is ultimately consumed by hardware, software largely determines how efficiently hardware components are utilized. Empirical studies demonstrate that software-level decisions can produce substantial differences in energy consumption: Pereira et al.~\cite{pereira2017energy} show that the choice of programming language alone can cause energy consumption to vary by more than an order of magnitude across implementations of the same algorithm, and Georgiou et al.~\cite{georgiou2022green} report significant energy differences between deep learning frameworks that perform equivalent computations. Consequently, improving code-level energy efficiency constitutes a direct and non-trivial lever for reducing the energy footprint of modern computing systems, independent of hardware choices.

Over the past decade, research and industry initiatives have advanced energy-efficient software development~\cite{hindle2012green,pinto2017energy}. Prior work has studied developer awareness~\cite{pang2015programmers,penzenstadler2014systematic}, tool support~\cite{oliveira2017study,weber2025famlem,fieni2024powerapi}, and recurring obstacles in energy-related development practices~\cite{pereira2017energy,dick2010model}. Despite this progress, energy considerations often remain secondary in day-to-day practice. 
%
Many developers recognize the technical relevance of energy consumption, yet consider it difficult to measure~\cite{khan2021measuring,damavsevivcius2013methods}, hard to optimize~\cite{kwon2013reducing,mittal2012empowering}, or challenging to justify under business contexts~\cite{funke2024experimental,wysocki2024don}. 
As a result, energy optimization rarely becomes a primary design or implementation objective in everyday software development workflows.

Despite numerous frameworks and tool proposals~\cite{georgiou2019software,di2017petra,aggarwal2015greenadvisor,jay2023experimental}, empirical evidence on how practitioners actually assess and act on energy consumption in day-to-day development remains limited. In this work, we aim to characterize the current state of practice among professional software developers with respect to energy assessment, and optimization, and we identify barriers practitioners perceive.
%
Specifically, we set out to answer the following overarching questions:
\emph{How do developers perceive the relevance of energy-efficient code?},
\emph{Which practices and indicators do they use to assess or estimate energy consumption?},
and \emph{Under what conditions do they consider energy optimization worthwhile?}

To address these questions, we conducted an online survey with 134 software developers across diverse domains and experience levels. The survey design builds on prior surveys~\cite{hoffmann2025sustainable,pathak2011bootstrapping,pinto2017energy,manotas2016empirical,pang2015programmers,pinto2014mining} while extending them with open-ended questions. We analyzed the qualitative responses via card sorting and category graph construction, and combined these results with quantitative analyses. This mixed-method design captures both broad patterns of practice and the rationale behind energy-related decisions, including points of disagreement within the community.

Our results indicate that explicit attention to energy-efficient software development is present in roughly one-third of the reported projects. Even when energy is not a stated project goal, about two-thirds of participants report optimizing properties that may influence energy use (e.g., runtime or CPU utilization) and recognize their potential impact on energy consumption. At the same time, approximately one-quarter of developers consider energy optimization not worth the effort in their context.

A central barrier is the perceived cost of accurate assessment: developers describe energy measurement as too difficult or too time-consuming when performed rigorously, suggesting demand for low-barrier measurement and feedback methods. Moreover, energy consumption is rarely captured systematically and regularly, even among developers who report caring about it, indicating structural barriers such as gaps in education, inadequate tooling, and organizational constraints.

When estimating energy consumption, developers often rely on proxy metrics such as runtime or CPU utilization because these are easier to obtain than energy measurements. Notably, \emph{electricity bills} are rated more favorably than metrics such as \emph{disk utilization} and \emph{RAM utilization}. Given that electricity bills are an aggregated and indirect outcome rather than a technical system metric, this preference suggests that some practitioners conceptualize energy primarily through economic abstractions rather than resource-level mechanisms, which has received limited attention in prior work yet.

Beyond using runtime as a proxy, many developers in our sample appear to rely on a mental model that largely equates runtime with energy consumption. While runtime is often a reasonable proxy in practice, there are edge cases where this relationship can weaken or break down~\cite{nakhkash2019analysis,mantovani2020performance}. In our data, only a small subset of developers identified such cases. Although our study was not designed or powered to assess individual differences, we observe preliminary indications that practitioners with more hands-on experience in managing software energy consumption may be better at recognizing situations where performance and energy diverge or even conflict.


Based on these findings, we derive three actionable recommendations: (i) integrate energy measurement and optimization into day-to-day development activities at the developer and team level; (ii) strengthen education to improve understanding of effort-benefit trade-offs and to reduce common misinterpretations about software energy consumption; and (iii) develop practical developer-oriented energy assessment and feedback tooling that lower the barriers for energy-aware programming.


In summary, this paper contributes:
\vspace{-1ex}
\begin{itemize}
    \item A mixed-method empirical characterization of current developer practices for assessing and optimizing code-level energy consumption, based on a survey of 134 professional software developers.
    \item Evidence that energy work is often indirect (embedded in other optimization goals) and largely confined to development activities rather than being systematically operationalized into production, including gaps in documentation and sustainable software engineering practices.
    \item A detailed account of practitioner reliance on proxy indicators for energy estimation, indicating an economic-oriented conceptualization of software energy consumption in practice.
    \item A publicly available replication package comprising the fully anonymized survey dataset and all analysis artifacts (including intermediate coding results and evaluation steps) to enable reproduction and support follow-up research~\footnote{Supplementary Web page: \supplMaterial.}.
\end{itemize}
\section{Related Work}
\label{rel:work}

Research on software energy consumption has expanded significantly in recent years, spanning measurement approaches, analytical methods, and optimization techniques. Early work focused on instrumentation and runtime measurement of energy usage in software systems~\cite{georgiou2019software, di2017petra, aggarwal2015greenadvisor, jay2023experimental, weber2025famlem}, alongside analytical approaches for modeling and estimating energy behavior~\cite{georgiou2022green, oliveira2019recommending, paniego2017analysis}. On the optimization side, established techniques range from compiler and system-level strategies to hardware-aware execution~\cite{cotes2017dynamic, leng2013gpuwattch}.

Beyond measurement and optimization, recent work has introduced formal models for quantifying the environmental impact of computation. Lannelongue et al.~\cite{lannelongue2021green} propose Green Algorithms, a generalizable framework for estimating the carbon footprint of computational workloads, enabling a reproducible and cross-comparable carbon footprint accounting across systems. Despite such methodological advances, their adoption and interpretation in practice remain limited.

%
%
%
Pinto et al.~\cite{pinto2014mining} analyzed developer discussions on Stack Overflow and found that questions about energy consumption are technically diverse and challenging, yet answers are often incomplete or imprecise. The study highlights two persistent issues: widespread misconceptions about software energy usage, such as conflating energy with power, or treating performance as a universal proxy for efficiency, and a lack of accessible measurement tools.

Further empirical studies corroborate these findings.
%
%
%
%
%
Pang et al.~\cite{pang2015programmers} surveyed 122 programmers and conducted follow-up interviews, revealing that most developers are unaware of the energy impact of their software, unfamiliar with measurement techniques, and uncertain about the causes of energy waste.
Similarly, Manotas et al.~\cite{manotas2016empirical} investigated practitioner perspectives at Microsoft and other large organizations, finding that developers rely primarily on performance profiling and indirect proxy metrics rather than direct energy measurements — and that energy efficiency requirements are often formulated implicitly, even when practitioners recognize these proxies as unreliable.

%
%
%
%
%
In the mobile domain, where energy constraints are particularly salient due to battery limitations, Pinto and Castor~\cite{pinto2017energy} studied how developers debug energy-related issues in open-source applications. The authors identify two key barriers: insufficient tool support (e.g., lack of integrated energy profilers) and an absence of structured knowledge sources such as guidelines and educational material. 
%
%
%
Pathak et al.~\cite{pathak2011bootstrapping} further show that energy bugs in mobile systems are diverse and often subtle, making them difficult to detect using traditional debugging approaches.

The broader research landscape has been reviewed from multiple angles. Hort et al.~\cite{hort2021survey} provide a comprehensive survey of performance optimization for mobile applications, noting that many such optimizations are indirectly motivated by energy constraints even when energy is not the explicit target. Verdecchia et al.~\cite{verdecchia2023systematic} offer a systematic review of Green AI, consolidating efforts across software engineering and machine learning. Their analysis emphasizes that, despite growing interest, the field remains fragmented, with limited convergence on standardized practices, metrics, and tooling---underscoring that energy-aware software engineering is still in a consolidation phase.

%
Recently, Majuntke et al.~\cite{hoffmann2025sustainable} surveyed practitioners and conducted expert interviews on the state of sustainable software development. Their findings point to persistent structural barriers: insufficient training, limited economic incentives, and organizational constraints that hinder the adoption of energy-efficient practices in everyday development.





\begin{table}
    \setlength{\extrarowheight}{2pt}
    \renewcommand{\arraystretch}{1.05}
    \centering
    \small
    \caption{Overview of samples of participants of previous surveys. \textbf{Refs in Table headings einabauen}}
    \label{tab:study_comparison}
    \begin{tabularx}{\textwidth}{L{1.4cm} Y Y Y Y Y}
        \toprule
        \textbf{Dimension} & \textbf{Pang et al.}~\cite{pang2015programmers} & \textbf{Manotas et al.}~\cite{manotas2016empirical} & \textbf{Pinto \& Castor}~\cite{pinto2017energy} & \textbf{Hoffmann et al.}~\cite{hoffmann2025sustainable} & \textbf{This study} \\
        \midrule

        Acquisition &
        Reddit &
        Company mailing lists and charts (ABB, Google, IBM, Microsoft) &
        Mobile open source developers with at least one commit to a mobile OSS project &
        Sustainability-focused networks (Bits\&Bäume, EcoCompute, ClimateActionTech, LinkedIn) &
        Reddit, LinkedIn, Mastodon, X, Bluesky, and professional networks \\

        \rowcolor{rowgray}
        Sample size &
        122 (survey) + 4 (interviews) &
        464 (survey) + 18 (interviews) &
        62 (survey) &
        109 (survey) + 27 (expert survey) + 5 (interviews) &
        134 (survey) \\

        Domain &
        Language-based only; no domain breakdown reported &
        Mobile, traditional PC, embedded, data center &
        Mobile open source only &
        No domain breakdown reported &
        Diverse; incl.\ mobile, embedded, IoT, backend, cloud, compiler \\

        \rowcolor{rowgray}
        Seniority &
        Collected but distribution not reported &
        Not reported &
        Reported &
        Not reported &
        Reported \\

        Geography &
        Not stated &
        USA-based companies &
        Not stated &
        Germany only &
        International \\

        \bottomrule
    \end{tabularx}
\end{table}

Table~\ref{tab:study_comparison} summarizes the key characteristics of the most closely related practitioner surveys. Across these surveys, the literature consistently identifies recurring challenges: limited developer awareness, insufficient tooling for energy measurement, reliance on imperfect proxy metrics, and a lack of systematic education and incentives. Although awareness is increasing in specific domains, such as mobile systems~\cite{pinto2017energy, pathak2011bootstrapping} and large-scale data-center environments~\cite{manotas2016empirical, hoffmann2025sustainable}, advances in formal quantification models~\cite{guldner2024development}, domain-specific optimization studies~\cite{pereira2017energy,rajput2024enhancing}, and cross-domain syntheses~\cite{pathania2025calculating,verdecchia2021green} indicate that the technical foundations are increasingly solid. Yet translating these advances into everyday development practice remains an open challenge.


Our study builds on these insights in several ways. First, we broaden the geographical and application domain scope: While prior surveys focus on specific companies~\cite{manotas2016empirical}, mobile open-source developers~\cite{pinto2017energy}, or a single country~\cite{hoffmann2025sustainable}, our sample spans diverse application domains and geographical regions. 
Second, rather than re-asking broad awareness questions already well covered in the literature (see Table~\ref{tab:researchquestion_to_surveyquestion}), we build directly on the pain points identified by prior work and target two persistent yet insufficiently understood aspects: (1) the incentives that motivate or discourage developers from writing energy-efficient software, and (2) misconceptions arising from the use of proxy metrics to estimate energy consumption. This allows us to move beyond mapping the problem and instead to probe \emph{why} energy-aware practices remain unsystematic despite growing technical and methodological support.

\section{Methodology}
\label{methodology}


\subsection{Research Questions}

We formulate our research questions guided by prior work~\cite{hoffmann2025sustainable,pathak2011bootstrapping,pinto2017energy,manotas2016empirical,pang2015programmers,pinto2014mining} on the state of energy-aware software development in practice. Overall, existing studies emphasize both the increasing importance of energy efficiency in software engineering and the persistent challenges of integrating energy considerations into development workflows. To capture the current state of practice, we formulate five research questions that examine developers’ perceptions, current practices, and future needs in energy-aware software engineering.

\par\addvspace{1ex}%
\noindent\RQ{1}: \emph{How do developers’ perceptions of the importance of software energy consumption align with actual prioritization and treatment in software projects?}\\
RQ\textsubscript{1} addresses whether energy consumption is perceived as a relevant and valuable criterion for decision-making, and the extent to which developers consider energy-related considerations to warrant attention and effort. Beyond perception, we are interested in how this perceived importance is reflected in practice by assessing the degree to which energy consumption is explicitly prioritized or addressed in software projects. By contrasting declared importance with observable project-level treatment, the question seeks to uncover mismatches between normative importance and practical relevance, thereby highlighting potential gaps between developers’ evaluative judgments and concrete engineering priorities.

\par\addvspace{1ex}%
\noindent\RQ{2}: \emph{How is software energy consumption assessed by practitioners amid challenges of integrating it into 
development workflows?}\\
Assessing software energy consumption remains a complex and effort-intensive task that has not yet seen widespread adoption in industry. Previous studies have emphasized the need for improved tooling and training to facilitate energy measurement in real-world projects~\cite{hoffmann2025sustainable, penzenstadler2014systematic}, where improved tooling refers to tools that are easier to use, more accurate, and more diverse in coverage, so that developers can select from a broader set of options suited to their specific application scenarios. RQ\textsubscript{2} aims at documenting the current state of practice, that is, how developers \emph{measure or estimate} energy consumption, what tools and techniques they use, and to what extent those assessment strategies are embedded into established development workflows.

\par\addvspace{1ex}%
\noindent\RQ{3}: \emph{What strategies and development practices are used to mitigate software energy consumption?}\\
RQ\textsubscript{3} focuses on the \emph{actions} developers take after identifying energy-related issues. We aim to understand how practitioners detect, analyze, and resolve energy inefficiencies in software systems. Specifically, we examine which techniques, tools, and development practices they use to detect energy problems and how they debug these problems. In particularly, we want to know how frequently they are applied, and how energy-aware software engineering practices are integrated into broader software development processes.

\par\addvspace{1ex}%
\noindent\RQ{4}: \emph{What mental models do practitioners use to relate software energy consumption to other metrics such as runtime performance?}\\
Measuring energy consumption is considerably more difficult than monitoring other non-functional properties, such as runtime performance or CPU utilization~\cite{pinto2014mining,weber2025famlem,manotas2016empirical}. Therefore, developers often rely on proxy metrics instead of direct energy measurements. While this approach is often reasonable, given that runtime and power jointly determine energy use, the correlation can vary or even invert depending on the execution context or system configuration. With RQ\textsubscript{4}, we investigate whether developers are aware of such trade-offs and how they reason about the relationship between energy consumption and other performance-related metrics in their daily work.

\par\addvspace{1ex}%
\noindent\RQ{5}: \emph{What strategies, resources, or incentives could support practitioners in better managing software energy use?}\\
Finally, we aim to identify persistent challenges and potential enablers for advancing energy-aware software engineering in practice. As developers are ultimately responsible for applying tools and methods in their daily work, we seek to understand which barriers hinder adoption and which forms of support, such as organizational incentives, educational resources, or improved tooling, could help foster more energy-aware software development practices.

\par\addvspace{1ex}%
Addressing these five research questions, we aim to provide a comprehensive understanding of the current state of practice in energy-aware software engineering, identify existing gaps between awareness and implementation, and highlight actionable pathways for improving the sustainability of software development.

\begin{figure}[t]
  \centering
  \includegraphics[width=\columnwidth]{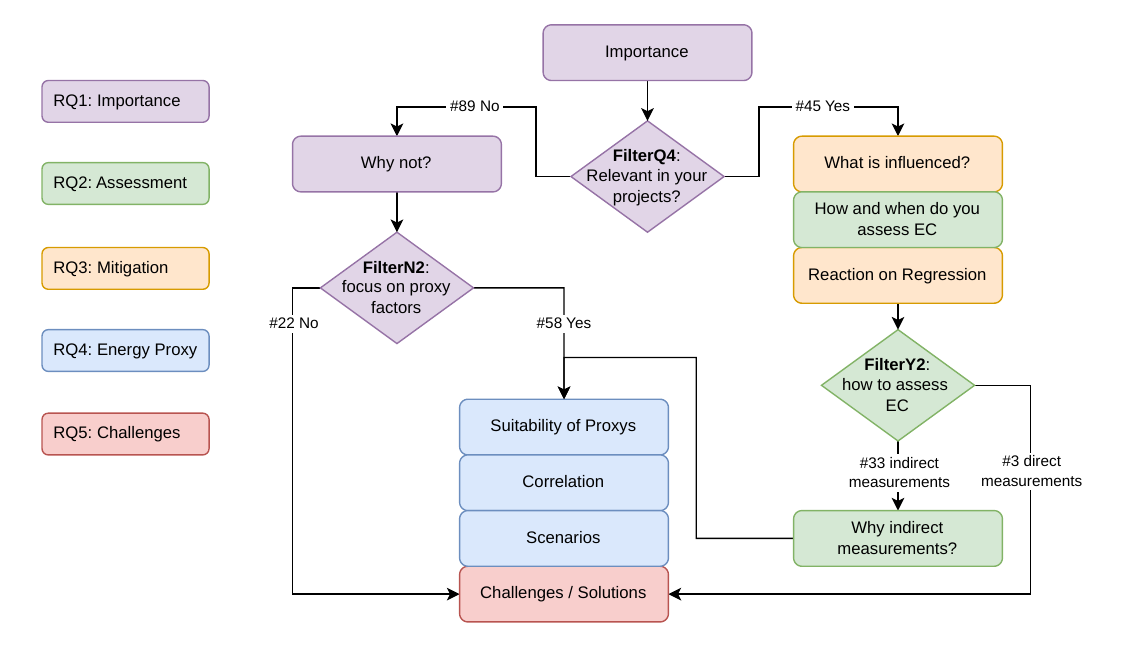}
  \caption{Overview of survey question topics, visualizing different paths through the study that participants take depending on their answers. Colors encode mapping of research questions to corresponding survey question blocks, EC stands for energy consumption. \textbf{FilterQ4}, \textbf{FilterN2}, and \textbf{FilterY2} describe the branching decision points of the survey based on which participants are guided through the survey and presented with branch-specific questions. \# indicates the count of participants who reached and responded to that stage of the survey; due to drop-out, these numbers do not sum to the total sample size at every branching point.}
  \Description{Overview of survey question topics, visualizing different paths through the study that participants follow depending on their answers. Colors encode mapping of research questions to corresponding survey question blocks, EC stands for energy consumption. \textbf{FilterQ4}, \textbf{FilterN2}, and \textbf{FilterY2} describe the branching decision points of the survey at which participants are guided through the survey and presented with branch-specific questions. \# indicates the count of participants who reached and responded to that stage of the survey; due to drop-out, these numbers do not sum to the total sample size at every branching point.}
  \label{fig:overview_of_question_topics}
\end{figure}

\subsection{Survey Questionnaire}

To answer our five research questions we designed a survey questionnaire with 22 closed and 6 open-ended questions (all questions and answer options are listed in Table~\ref{tab:researchquestion_to_surveyquestion}). The survey questions focused on participants' experiences and opinions regarding the importance, measurement, and optimization of energy consumption throughout the software engineering life cycle. The questionnaire is structured around our research questions, covering five areas: (RQ1)~perceived importance of software energy consumption for various stakeholders, (RQ2)~energy consumption assessment, (RQ3)~mitigation of energy bugs, (RQ4)~use of proxy metrics, and (RQ5)~perceived challenges and promising solutions. 
Fig.~\ref{fig:overview_of_question_topics} illustrates the order in which questions were presented. The diamond-shaped nodes (\textbf{FilterQ4}, \textbf{FilterN2}, and \textbf{FilterY2}) represent the branching decision points of the survey, at which participants are guided through the survey and presented with branch-specific questions depending on their previous answers. This approach kept the questionnaire concise and avoided presenting possibly irrelevant questions. For example, participants who indicated that energy consumption is not relevant to their projects were not asked about its impact or measurement. Instead, they were asked why it is not relevant to their projects.

As shown in Table~\ref{tab:researchquestion_to_surveyquestion}, a subset of our survey questions builds on questions previously asked in related practitioner surveys~\cite{pinto2017energy, pang2015programmers, manotas2016empirical, hoffmann2025sustainable}. However, nine questions are novel contributions of this survey. This includes questions in the areas of energy consumption assessment, mitigation of energy bugs, and the use of proxy metrics, as shown in Table~\ref{tab:researchquestion_to_surveyquestion} and Table~\ref{tab:researchquestion_to_surveyquestion_scenarios}.

\begin{center}
    \setlength{\extrarowheight}{2pt}
    \renewcommand{\arraystretch}{1.05}
    \small
    
    \begin{tabularx}{\textwidth}{L{0.2cm} Y L{5.8cm} L{1.8cm}}

        \caption{Mapping of survey questions to corresponding research questions.}\\
        \toprule
        \textbf{RQ} & \textbf{Question} & \textbf{Answer options} & \textbf{Related Work} \\
        \midrule
        \endfirsthead
        \caption{Mapping of survey questions to corresponding research questions. \textit{continued}}\\
        \toprule
        \textbf{RQ} & \textbf{Question} & \textbf{Answer options} & \textbf{Related Work} \\
        \midrule
        \endhead

        \midrule
        \multicolumn{4}{r}{\textit{Continued on next page}}\\
        \endfoot

        \bottomrule
        \endlastfoot

        \label{tab:researchquestion_to_surveyquestion}

        1 & 
        How important is the energy consumption of your software product for ...  \par
                \par\hspace*{4em}...you? \par\hspace*{4em}...your colleagues? \par\hspace*{4em}...your company? \par\hspace*{4em}...your customers? 
        & \par\opt 1 - unimportant \par\opt 2 \par\opt 3 \par\opt 4 \par\opt 5 - important 
        & \cite{pang2015programmers, manotas2016empirical} \\

        \rowcolor{rowgray}
        1 & 
        Is the energy consumption of your software product a practical issue in your projects that influences decisions and actions? & 
        \opt Yes, energy consumption is a practical issue and we assess the energy consumption of our software. 
        \par\opt No, energy consumption is not a practical issue, even if we focus on other properties that, in turn, may affect energy consumption. 
        & \cite{manotas2016empirical, pinto2017energy} \\

        1 & 
        In your opinion, is it worth the effort to reduce energy consumption by optimizing software? Please elaborate. & 
        free text 
        & \cite{manotas2016empirical} \\

        \rowcolor{rowgray}
        2 & 
        Do you focus on other factors that might influence energy consumption? & 
        \opt yes 
        \par\opt no 
        & \cite{manotas2016empirical} \\

        2 & 
        How do you assess the energy consumption of your software? &
        \opt Entire system at the power plug 
        \par\opt RAPL using on-board sensors
        \par\opt runtime
        \par\opt CPU utilization
        \par\opt RAM utilization
        \par\opt disk utilization
        \par\opt network utilization
        \par\opt battery usage
        \par\opt cloud provider costs
        \par\opt electricity bills
        \par\opt based on customer feedback
        \par\opt based on feeling or intuition
        \par\opt other 
        & \cite{manotas2016empirical, hoffmann2025sustainable} \\

        \rowcolor{rowgray}
        2 & 
        What are the reasons for you and your company that you do not measure energy consumption directly? \& Given your answers, you assess energy consumption directly and indirectly. Why do you use indirect measurements additionally to direct power measurements? &
        free text 
        & \cite{manotas2016empirical} \\

        2 & 
        When do you assess energy consumption of your software system?  &
        \par\opt after each commit
        \par\opt after each feature completion
        \par\opt after each release/deployment
        \par\opt irregularly (based on intuition)
        \par\opt irregularly (based on feedback or request)
        \par\opt never
        \par\opt other 
        & \cite{manotas2016empirical} \\

        \rowcolor{rowgray}
        2 & 
        On which of these levels do you assess energy consumption? &
        \par\opt code level
        \par\opt feature level
        \par\opt module level
        \par\opt system level
        \par\opt development Process
        \par\opt other 
        & \textit{novel} \\

        2 & 
        Do you automate the assessment of energy consumption (e.g. within a CI/CD pipeline)? &
        \opt yes
        \par\opt no
        \par\opt no answer 
        & \textit{novel} \\

        \rowcolor{rowgray}
        3 & Which decisions in your daily practice are influenced by considerations regarding energy consumption of your software? & 
        \par\opt decisions on specific hardware components 
        \par\opt decisions on specific frameworks/libraries 
        \par\opt decisions on Software design and construction 
        \par\opt decisions on source-code level 
        \par\opt other 
        & \cite{manotas2016empirical} \\

        3 & How often do you make decisions based on energy consumption? & 
        \par\opt regularly 
        \par\opt sometimes 
        \par\opt never 
        & \cite{manotas2016empirical} \\

        \rowcolor{rowgray}
        3 & What actions do you usually take when the energy consumption of your software increases? & 
        \par\opt documenting the energy regression 
        \par\opt analyzing the root-cause of the energy regression 
        \par\opt fixing identified causes for the energy regression 
        \par\opt we ignore it 
        \par\opt I don't know 
        \par\opt other 
        & \textit{novel} \\

        3 & Is the debugging process for energy problems different from the debugging process for performance problems or functional errors? & 
        \par\opt yes 
        \par\opt no 
        \par\opt I don't know 
        \par\opt no answer 
        & \textit{novel} \\

        \rowcolor{rowgray}
        3 & Please elaborate on similarities and differences between debugging energy bugs and other bugs: & free text 
        & \textit{novel} \\

        4 & 
        How suitable are the following metrics to act as a proxy for energy consumption of software in your opinion? 
        \par\hspace*{4em} Runtime \par\hspace*{4em} CPU utilization \par\hspace*{4em} RAM utilization \par\hspace*{4em} Disk utilization \par\hspace*{4em} Network utilization \par\hspace*{4em} Battery usage \par\hspace*{4em} Cloud provider costs \par\hspace*{4em} Electricity bills & 
        \opt highly suitable 
        \par\opt somewhat suitable 
        \par\opt somewhat unsuitable 
        \par\opt completely unsuitable 
        \par\opt not applicable 
        & \textit{novel} \\

        \rowcolor{rowgray}
        4 & Do you think there is a correlation between runtime and energy consumption? & free text 
        & \textit{novel} \\

        5 & Reflecting on your previous answers, what challenges do you face with energy consumption in your software projects? & free text 
        & \cite{pang2015programmers, manotas2016empirical, hoffmann2025sustainable, pinto2017energy} \\

        \rowcolor{rowgray}
        5 & What do you think needs to be done to address the challenges of energy consumption? & free text 
        & \cite{manotas2016empirical, hoffmann2025sustainable} \\

        \bottomrule
    \end{tabularx}
\end{center}

\subsection{Correlation Scenarios}\label{sec:scenarios}
To assess developers' judgments about the relationship between energy consumption and proxy metrics, such as runtime performance, we employ scenario-based questions grounded in real-world software systems (exact scenario descriptions are shown in Table~\ref{tab:researchquestion_to_surveyquestion_scenarios}). Scenarios allow us to elicit participants' assessments of whether and how energy consumption and performance are related in concrete execution contexts. The selected scenarios are derived from prior empirical work that explicitly investigates the energy-performance relationship in practice~\cite{yazdanbakhsh2016axbench,weber2023twins,cadenelli2017accelerating,wu2024abakus,rucci2015energy}. To the best of our knowledge, using scenario-based questions to empirically probe developers' mental models of the energy-performance relationship is a novel contribution of this survey; prior practitioner surveys~\cite{pang2015programmers, manotas2016empirical, hoffmann2025sustainable} have examined awareness and practices but have not employed concrete execution scenarios to assess whether proxy-based reasoning translates into correct expectations.
Participants are not expected to answer all scenarios correctly on an individual level. Reasoning about energy consumption requires substantial knowledge of the setup of the software system, including CPU scheduling, memory behavior, and I/O effects. Instead, our analysis focuses on whether the collective judgments of the participant cohort exhibit a tendency toward the correct correlation mode for each scenario.

\textit{Strong positive correlation.}
This scenario describes a video encoding task executed single-threaded on a CPU. Video encoding is a compute-intensive workload with minimal I/O and memory contention. For such CPU-bound tasks, prior work reports a strong and approximately linear positive correlation between runtime performance and energy consumption~\cite{yazdanbakhsh2016axbench,weber2023twins}. Faster execution reduces the total active CPU time and, consequently, overall energy consumption. The absence of parallelism further limits scheduling and context-switching overhead, reinforcing this strong positive correlation.

\textit{Moderate positive correlation.}
This scenario extends the previous one by enabling multi-threaded execution across multiple CPU cores. While the correlation between performance and energy consumption remains positive, it becomes weaker. Parallel execution reduces runtime substantially. However, energy consumption does not decrease proportionally. Context switches, synchronization, and parallel runtime overhead introduce periods in which energy is consumed without productive computation. As a result, using additional CPU cores improves performance more strongly than it reduces energy consumption, altering the energy–performance ratio compared to single-threaded execution~\cite{weber2023twins}.

\textit{Negative correlation.}
This scenario represents a data-intensive workload executed on multi-core hardware, involving frequent memory accesses~\cite{cadenelli2017accelerating, wu2024abakus, rucci2015energy}. Although parallelization reduces runtime, prior measurements show that energy consumption can increase, for example due to memory activity and coordination overhead between CPU cores~\cite{rucci2015energy}. In this case, performance improvements come at the cost of higher energy usage, resulting in a negative correlation between energy consumption and performance (i.e., shorter execution time). This scenario highlights the role of memory behavior and multi-core interactions in shaping the energy–performance relationship.

\textit{No correlation.}
The final scenario considers a database system in which execution time is essentially constant across configurations, whereas energy consumption exhibits substantial variability. Weber et al.~\cite{weber2023twins} report energy deviations of up to 25~\% under identical execution times, attributable to configuration choices that alter access patterns to storage and memory. Related work likewise shows that configurations across diverse domains (e.g., multi-core environments, mobile processors, and DNNs) can markedly affect energy consumption without inducing measurable performance changes~\cite{li2016evaluating,chetsa2014exploiting,dzhagaryan2014impact}. Accordingly, this scenario captures a setting in which energy consumption and performance are largely uncorrelated.

\begin{table}
    \setlength{\extrarowheight}{2pt}
    \renewcommand{\arraystretch}{1.05}
    \centering
    \small
    \caption{Scenario descriptions and corresponding answer options (correct answer shown in bold) for RQ\textsubscript{4}, with corresponding scenario ID's. All four scenarios are novel contributions of this survey.}
    \label{tab:researchquestion_to_surveyquestion_scenarios}
    
    \begin{tabularx}{\textwidth}{C{1.2cm} Y L{7cm} L{1.8cm}}
        \toprule
        \textbf{ID} & \textbf{Scenario} & \textbf{Question} & \textbf{Related Work} \\
        \midrule
        
        \rowcolor{rowgray}
        \centering\rowdesc{strong positive\\correlation}
        &
        A video encoder program is running single-threaded. When encoding the same video twice with different encoder settings, the execution time for one encoding is twice as long as the other.
        &
        I expect that the encoding of the video with longer runtime \dots \begin{itemize}[label=\opt,leftmargin=*,labelsep=.6em,itemsep=0pt,nosep,topsep=.2em,parsep=0pt]
          \item \textbf{consumes double energy}
          \item consumes more energy, but not double
          \item consumes less energy, but not double
          \item consumes half energy
          \item shows no systematic change in the energy consumption
        \end{itemize}
        & \textit{novel} \\

        \centering\rowdesc{moderate positive\\correlation}
        &
        A video encoder program is running multi-threaded. When encoding the same video twice with different encoder settings, the execution time for one encoding is twice as long as the other.
        &
        I expect that the encoding of the video with longer runtime \dots \begin{itemize}[label=\opt,leftmargin=*,labelsep=.6em,itemsep=0pt,nosep,topsep=.2em,parsep=0pt]
          \item consumes double energy
          \item \textbf{consumes more energy, but not double}
          \item consumes less energy, but not double
          \item consumes half energy
          \item shows no systematic change in the energy consumption
        \end{itemize}
        & \textit{novel} \\

        \rowcolor{rowgray}
        \centering\rowdesc{negative\\correlation}
        &
        A software program is designed to align sequences of nucleotides from several gigabytes of DNA input data. This is a data-intensive task. By leveraging multi-core hardware, users were able to significantly reduce the execution time to perform the task.\vspace{\baselineskip}
        &
        I expect energy consumption to \dots
        \begin{itemize}[label=\opt,leftmargin=*,labelsep=.6em,itemsep=0pt,nosep,topsep=.2em,parsep=0pt]
            \item be also significantly reduced
            \item be also reduced
            \item \textbf{be increased}
            \item \textbf{be significantly increased}
            \item to not systematically change
        \end{itemize}
        & \textit{novel} \\

        \centering\rowdesc{no\\correlation}
        &
        A software team evaluates different configurations of a database. The team identifies a set of configurations under which the database has the same runtime. Choose the option best describing your expectation.\vspace{\baselineskip}
        &
        I expect that all configurations of the identified set consume \dots
        \begin{itemize}[label=\opt,leftmargin=*,labelsep=.6em,itemsep=0pt,nosep,topsep=.2em,parsep=0pt]
            \item the same amount of energy
            \item a similar amount of energy, with small variations
            \item \textbf{differing amounts of energy, even though they have the same runtime}
        \end{itemize}
        & \textit{novel} \\
        
        \bottomrule
    \end{tabularx}
\end{table}


\subsection{Participants}
\paragraph{Acquisition}
To maximize visibility and participation, we recruited participants through a set of online communities and professional channels that are likely to include practitioners with hands-on experience in performance and systems optimization, and thus potentially in energy-related concerns. On \emph{Reddit}, we targeted relevant subreddits (e.g., \emph{C++} and \emph{Rust}) and contacted moderators in advance to obtain permission before posting the study announcement. Beyond Reddit, we promoted the survey through \emph{Mastodon}, \emph{X (formerly Twitter)}, \emph{LinkedIn} (with support from the \emph{Bundesverband Green Software} in Germany), \emph{Bluesky}, and \emph{Medium}. We prepared platform-specific promotional texts\footnote{Promotional texts are available on the supplementary Web page: \supplMaterial.} to match the tone and conventions of each community. In addition, we approached companies in our professional networks via email and encouraged them to share the invitation internally with employees and collaborators. While this strategy was not designed to yield a representative sample of the entire software engineering population, it intentionally increases the likelihood of recruiting respondents with an expertise in energy-aware software engineering practices.

\paragraph{Demographics}


A total of 175 participants started the survey. To ensure data relevance, we included only those respondents who answered survey question \emph{Q4: ``Is the energy consumption of your software product a practical issue in your projects that influences decisions and actions?''}, yielding a final sample of 134 participants. This question determines the branching path through subsequent parts of the questionnaire and enables inclusion of participants into the evaluation of RQ\textsubscript{1} (see Fig~\ref{fig:overview_of_question_topics}). Since participants could leave the survey at any point, the number of respondents contributing to the different branches of the survey varies and does not always sum to 134. Throughout the paper, we report the number of participants who reached and responded to the respective question; 103 participants completed the survey.

The set of participants represents a broad spectrum of career stages and works in a variety of software domains, including mobile, embedded, and IoT systems. They have experience working across different layers of software systems, such as backend, frontend, enterprise, Web, cloud, security, and compiler development, providing a diverse cross-section of perspectives from contemporary software engineering practice.

\subsection{Analysis}

We employed open card sorting~\cite{H14,Z16} to extract common themes and arguments from the open-ended responses of the participants. Open card sorting is a qualitative analysis technique in which arguments are freely grouped into categories that are defined in the process, without a predetermined taxonomy~\cite{H14,Z16}. This makes it particularly suitable for exploratory analysis of free-text responses, where the goal is to inductively derive a thematic structure from the data rather than to validate an existing one. The analysis was conducted collaboratively by two members of the author team over multiple sessions (each lasting 3–5 hours). For each answer, we generated one card, including a question identifier, a participant identifier, and the participant's answer. Starting with the first card, each of the two authors independently read the response. Then, they discussed and jointly assigned it to a category. Whenever an answer did not adequately fit into an existing category, a new category was created. If a response covered multiple categories, the card was physically split, and each part was assigned separately. After sorting all responses for a question, all categories and assignments were reviewed again to ensure internal consistency and coherence.

We compared responses across questions to deepen our understanding of the data. In particular, we contrasted arguments from developers who indicated that energy consumption already plays a role in their projects with those who reported that it does not. This comparison allowed us to explore how perceptions, practices, and challenges differ depending on the relevance of energy considerations in participants’ professional contexts.

\paragraph{Correlation Scenario Judgement Accuracy.}

Finally, we analyze the judgment accuracy of the participants regarding the question whether performance is a reliable proxy for energy consumption for each scenario. We operationalize the accuracy of developer's judgments using a ternary metric. A response is classified as \textit{correct} (0) if it matches the ground truth exactly; as \textit{partially correct} (1) if it reflects the right direction but wrong strength (e.g., selecting \textit{moderate positive} when the correct answer is \textit{perfect positive}); and as \textit{incorrect} (2) if it is neither directionally nor substantively aligned with the ground truth. This encoding preserves the meaningful distinction between a directionally plausible near-miss and a fully incorrect response, while remaining conservative enough to be applicable given the small group sizes that result from stratification by demographic variables. We report medians and interquartile ranges (IQRs) as descriptive statistics throughout, consistent with the ordinal nature of the outcome.

We use non-parametric significance tests, which is most suitable for ordinal outcomes and small, unequal group sizes. To compare \textit{expert vs.\ non-expert} developers, we apply the \textbf{Mann-Whitney U test} for two independent groups. For \textit{seniority} (Junior / Senior / Lead) and \textit{domain} (Mobile, IoT, Web Development, Infrastructure, Enterprise), we apply the \textbf{Kruskal-Wallis H test}; where results are significant, we follow up with \textbf{Dunn's post-hoc test} (Bonferroni corrected) to identify differing group pairs. For \textit{years of energy-aware development}, treated as continuous, we compute \textbf{Spearman's rank correlation} ($\rho$). All tests are evaluated at $\alpha = 0.05$.

\section{Results}
\label{study:Results}

For each research question, we first present participants' answers to closed questions. Next, we describe the results of the qualitative analysis by introducing all final themes.


\subsection{\RQ{1}: How do developers’ perceptions of the importance of software energy consumption align with actual prioritization and treatment in software projects?}


%


\subsubsection{Quantitative insights: Perceived and actual importance}

\begin{figure*}[t!]
    \centering
    \begin{subfigure}[t]{0.70\textwidth}
        \centering
        \includegraphics[width=\columnwidth]{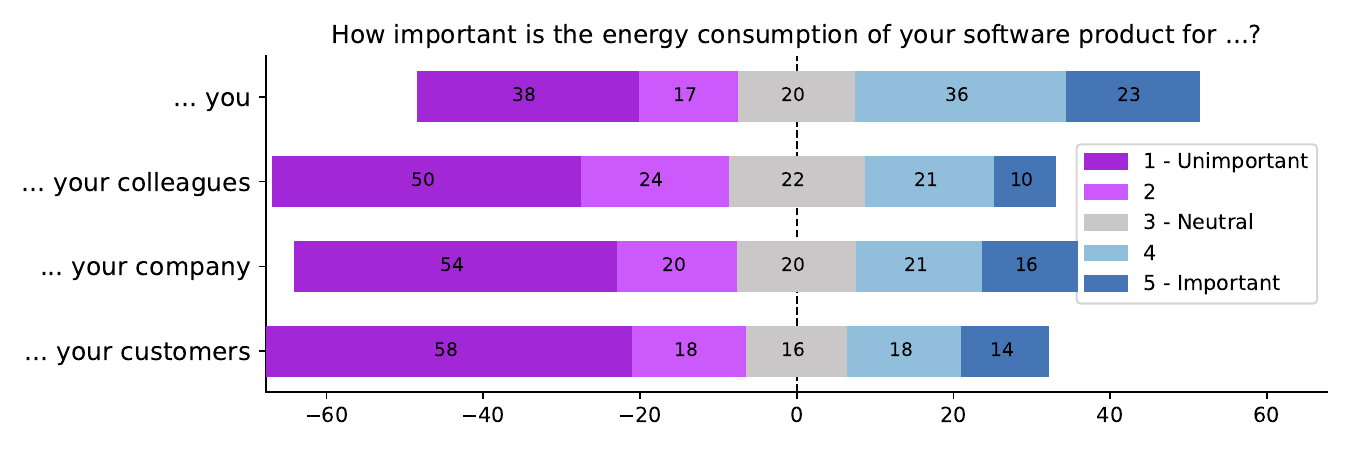}
        \caption{Developers perception of importance of software energy consumption}
        \label{fig:importance_perceived}
    \end{subfigure}%
    \hfill%
    \begin{subfigure}[t]{0.28\textwidth}
        \centering
        \includegraphics[width=\columnwidth]{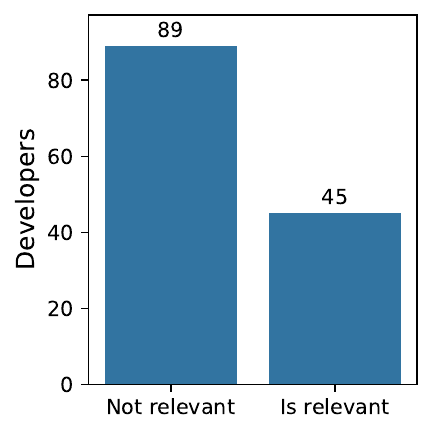}
        \caption{Actual relevance in developers' projects.}
        \label{fig:importance_actual}
    \end{subfigure}
    \caption{Importance of software energy consumption.}
    \Description{Image showing the perceived and actual importance of software energy consumption for developers.}
    \label{fig:importance}
\end{figure*}


Fig.~\ref{fig:importance_perceived} shows the distribution of participants' responses regarding the importance of energy consumption for themselves, their colleagues, and other stakeholders. We find that 59 participants consider energy consumption important to themselves, while 55 consider it unimportant, and 20 have a neutral stance.
In contrast, the perceived importance of energy consumption for other stakeholders is noticeably lower: most participants believe it is unimportant to their colleagues (74), their company (74), and their customers (76). So, conversely, far fewer consider it important for colleagues (21), their company (27), or customers (32).
This indicates a substantial group of developers in our sample who view energy consumption as more important personally than others. This pattern aligns with prior findings that developers assess their own environmental awareness more positively than that of others, as found by Schneider and Betz~\cite{schneider2022transformation2}.


To assess the practical relevance of energy-aware software development, we asked developers to reflect on their current projects and to what extent software energy consumption influences their decisions and actions. Specifically, we were interested in whether energy consumption constitutes an explicit concern in development and, if not, whether developers prioritize other factors that implicitly affect energy use.
Among the 134 participants, 45 reported that software energy consumption already plays a recognizable and direct role in their development decisions (as shown in Fig.~\ref{fig:importance_actual}). The remaining 89 participants indicated that energy consumption is not an explicit priority. Within this group, two distinct patterns emerged. For 22 participants, energy consumption is considered irrelevant due to the application domain or project context, whereas 58 participants reported focusing on technical attributes such as performance, battery lifetime, or memory consumption, which indirectly shape the energy behavior of the software.
These findings reveal a clear distinction between explicit engagement at which energy consumption is treated as a first-class requirement, and implicit engagement at which energy-related outcomes arise as a byproduct of optimizing proxy properties. Taken together, these results indicate that in the majority of projects represented in our study (103 of 134), software energy consumption is affected by development decisions and practices.
%
\begin{figure}[t]
  \centering
  \includegraphics[width=\columnwidth]{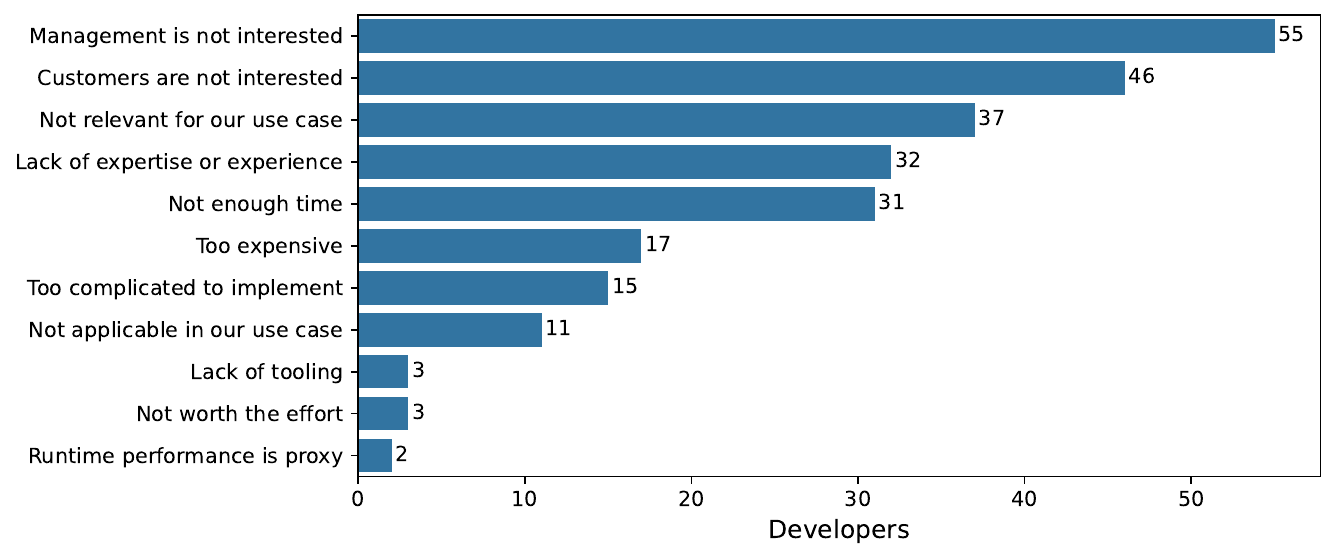}
  \vspace{-6ex}
  \caption{Developers' arguments for not considering software energy consumption.}
  \Description{Image showing developers' arguments for not considering software energy consumption.}
  \label{fig:no_energy_assessment_arguments}
\end{figure}

We asked the 89 participants who expressed that energy consumption is not an explicit priority for reasons why energy consumption is not explicitly addressed. 
Fig.~\ref{fig:no_energy_assessment_arguments} summarizes their answers. Most common reasons include a lack of management interest (55), a lack of customer demand (46), and irrelevance for the current use case (46). Developers also emphasized a lack of expertise (32), lack of time (31), or high perceived complexity (16). Across these responses, two patterns stand out: organizational priorities (management, customer requirements) heavily influence whether energy is of interest, as well as technical barriers and missing education, which continues to limit the adoption of energy assessment tools and practices.
Building on these results, we analyze how developers assess software energy consumption and how they address energy regressions in practice in RQ\textsubscript{2} and RQ\textsubscript{3}.

\subsubsection{Open questions: Implementation in practice}

\begin{figure}[t]
  \centering
  \includegraphics[width=\columnwidth]{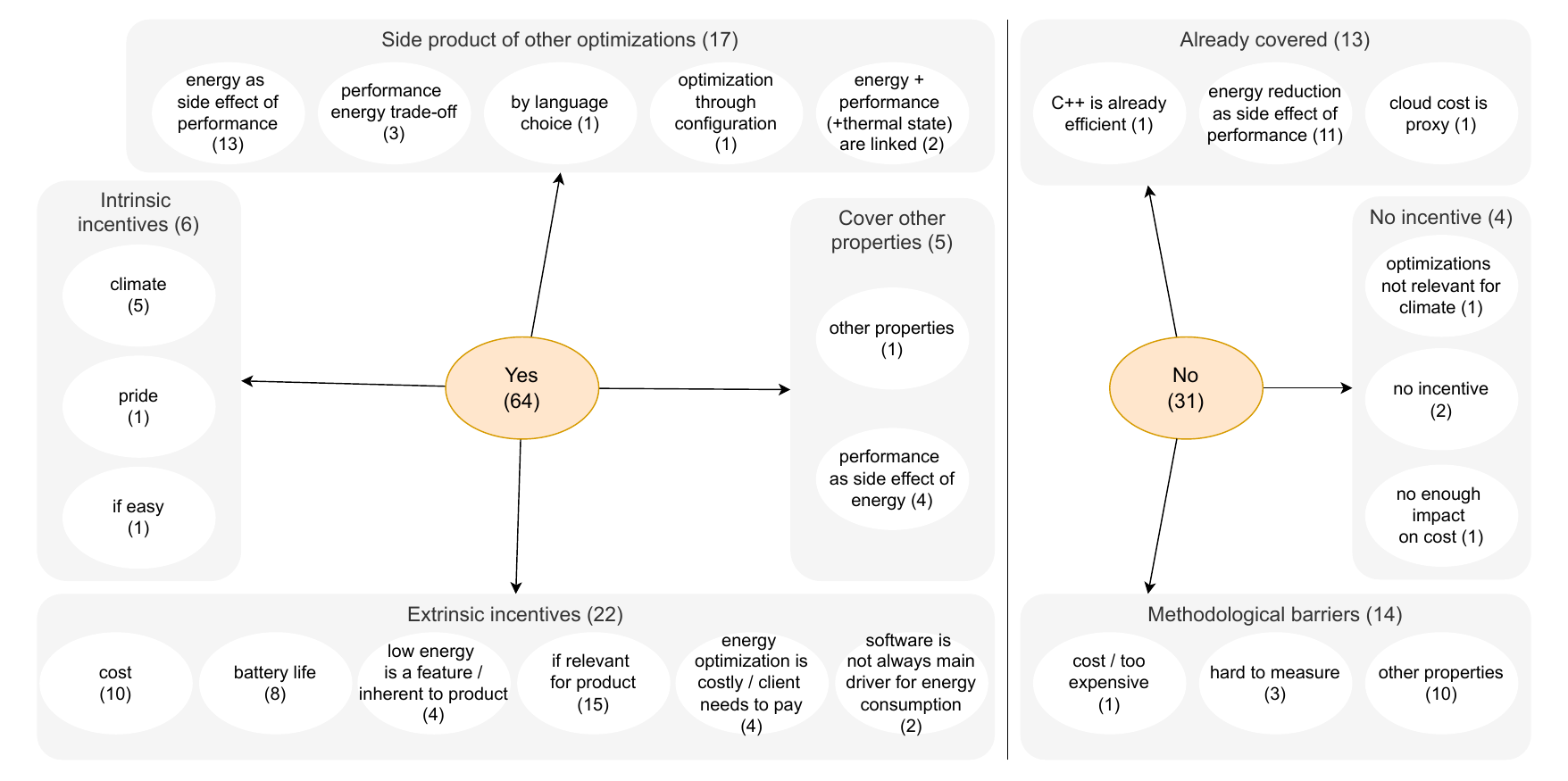}
  \caption{Developer's arguments on whether optimizing software energy is worthwhile. Survey question: ``From your perspective, is the effort to reduce energy consumption through software optimization worthwhile? Please explain.''}
  \Description{Image showing developer's arguments on whether optimizing software energy is worthwhile. Survey question: ``From your perspective, is the effort to reduce energy consumption through software optimization worthwhile? Please explain.''}
  \label{fig:effort_arguments}
\end{figure}


Participants gave open-ended explanations regarding whether they consider energy optimization worthwhile (see Fig.~\ref{fig:effort_arguments}). Across all responses, 64 support energy optimization and 31 argue against it. We clustered the arguments into seven groups, which we cover in detail next, starting with arguments in favor of energy optimization.

\textit{Energy is optimized as a side product of other optimizations (17).}
A major supportive theme emphasizes that energy optimization naturally accompanies optimization of other software properties. One participant noted that reducing power use often appears as \simplequote{a side benefit of reducing server load and response times}{p33}\footnote{Participant identifiers (e.g., \textit{$_{p33}$}) refer to anonymized participant IDs. They correspond to the identifiers in the results table available on the supplementary web page: \url{https://github.com/AI-4-SE/Attitudes-and-Practices-Towards-Optimizing-Software-Energy-Consumption}.}. Another pointed to language choice, stating a preference for \simplequote{efficient languages}{p114} viewed as inherently less wasteful.

\textit{Energy optimization to optimize other properties (5).}
A second theme reverses the dependency. Some developers see energy optimization as a direct means of improving other qualities. One commented that improving energy \simplequote{leads to the software running more efficiently and places fewer demands on the hardware}{p212}, where another emphasized that greater efficiency \simplequote{runs faster, so that's already a benefit}{p135}.

\textit{Energy optimization due to extrinsic (22) and intrinsic (6) incentives.}
Additional positive arguments reflect extrinsic motivations. Some highlight cost savings, stating that lower energy consumption \simplequote{reduces the electricity bill}{p110}. Others point to battery-constrained environments, noting that efficiency \simplequote{extends battery life of mobile phones and embedded devices}{p95}. Developers working in specialized domains further stress product relevance. For instance, one participant in the hearing-aid domain described it as central because \simplequote{energy consumption has a direct impact on the usage duration between charging the battery}{p42}.
Beyond these domain-specific arguments, respondents also referenced internal incentives tied to climate responsibility and professional identity. Some emphasized the environmental dimension, arguing that \simplequote{saving on energy consumption can basically never be a bad thing because high energy consumption costs money and hurts the climate}{p199}. Others framed energy efficiency as a matter of personal and professional ethos, noting that \simplequote{I generally feel it's part of my responsibility as an engineer}{p141}. Finally, participants also conveyed a pragmatic stance, expressing willingness to pursue energy-efficient solutions when the required effort remains low: \simplequote{What can be implemented easily [...] should be done as well, even if customers do not ask for it.}{p19}.

\textit{Energy is not optimized, due to methodological barriers (14), lack of incentives (4), or energy consumption is already managed (13).}
Arguments against prioritizing energy optimization fall into three clusters. Some highlight methodological barriers, stating that \simplequote{there are no stable metrics available and no established methods known that allow the energy consumption of software to be reduced in a traceable way.}{p15}. Others point to limited incentives, summarizing the issue as \simplequote{I don't pay the bills}{p207}. A final group argues energy consumption is already indirectly managed, particularly in cloud-based development. One developer expressed that with modern cloud platforms \simplequote{energy already correlates with spending. Not sure why I should think beyond that for backend software}{p91}.

\subsubsection{Discussion}

We observe a strong contrast between developers' self-evaluations and their evaluations of colleagues. Notably, developers' assessment of others aligns well with what is reported for current project practice. That is, energy consumption is not relevant in roughly two thirds of projects and relevant in about one third, as shown in Fig.~\ref{fig:importance}. In contrast, developers' stated personal importance of software energy consumption does not map onto the relevance of energy consumption in their projects. Possible reasons are: (1) developers may hold a higher internal standard than what they ultimately implement under real-world constraints, or (2) we may be observing a self-enhancement bias. Self-enhancement bias describes individuals' tendency to maintain unrealistically positive self-views~\cite{dufner2019self}. In our setting, this would mean that developers perceive themselves as caring more about software energy consumption than their peers.

These findings are consistent with the broader picture obtained from prior practitioner surveys. Pang et al.~\cite{pang2015programmers} found that only 18\,\% of programmers took energy consumption into account when developing software, and Majuntke et al.~\cite{hoffmann2025sustainable} report a near-identical figure a decade later, with only 17\,\% of respondents having ever applied green coding practices. Manotas et al.~\cite{manotas2016empirical} similarly found that the majority of practitioners have have never or rarely energy requirements, while Pinto and Castor~\cite{pinto2017energy} observed that most developers lack the knowledge and tools to act on energy concerns even when they recognize them. 
\added{Our finding that 45 of 134 participants (34\,\%) explicitly consider energy consumption in their projects is not directly comparable to these figures, as the studies differ substantially in population, sampling, and operationalization of energy engagement. We therefore do not interpret the higher percentage as evidence of a quantitative increase. Nevertheless, taken together, these independently conducted studies, spanning roughly a decade, consistently place practitioner engagement with energy concerns in a low-to-moderate minority, with no study reporting majority engagement. This convergent pattern across heterogeneous methodologies lends qualitative support to the view that energy awareness remains an emerging rather than mainstream concern, even as it appears to be gradually gaining traction.}

The distinction between explicit and implicit engagement with energy consumption is central to understanding how energy considerations manifest in practice. While only a minority works in environments where energy is an explicit requirement, a much larger group repeatedly utilizes properties such as performance, battery consumption, and memory footprint, which have direct consequences for energy use. Under this broader framing, 103 of 134 participants (77\,\%) in our study influence software energy consumption through their development decisions, a figure substantially higher than any prior finding. This confirms that energy considerations often materialize indirectly, embedded within other design trade-offs. Yet prior research rarely addresses the implications of such proxy-driven interactions. Our findings indicate that overlooking such indirect measures understates the degree to which developers already influence energy efficiency.

The qualitative arguments explain why developers diverge in their views. Supportive arguments tend to emphasize synergies: optimizing energy either emerges naturally from established performance practices or drives improvements in other system qualities. These perspectives position energy optimization as aligned with existing workflows rather than as an additional development task. Negative arguments stem from structural disincentives and methodological uncertainty. Developers struggle with unclear measurement practices and absent organizational incentives, particularly when energy costs are abstracted away through cloud infrastructure or absorbed by users.


\added{Together, these insights portray a development ecosystem in which energy consumption is recognized as relevant by 44\,\% of developers (59 of 134). This skew toward ``unimportant" suggests that perceived importance alone is not a reliable predictor of practice: developers who see energy as synergistic with established workflows are more likely to act on it, regardless of whether they explicitly rate it ``important," while those who perceive misaligned incentives or insufficient methodological support deprioritize it even when they consider it relevant. Strengthening measurement tooling, clarifying organizational priorities, and making implicit energy implications more transparent could bridge the gap between awareness and consistent practice.}


\subsection{\RQ{2}: How is software energy consumption assessed by practitioners amid challenges of integrating it into development workflows?}

\subsubsection{Quantitative insights: Assessment of software energy consumption}

\begin{figure}[t]
  \centering
  \includegraphics[width=\columnwidth]{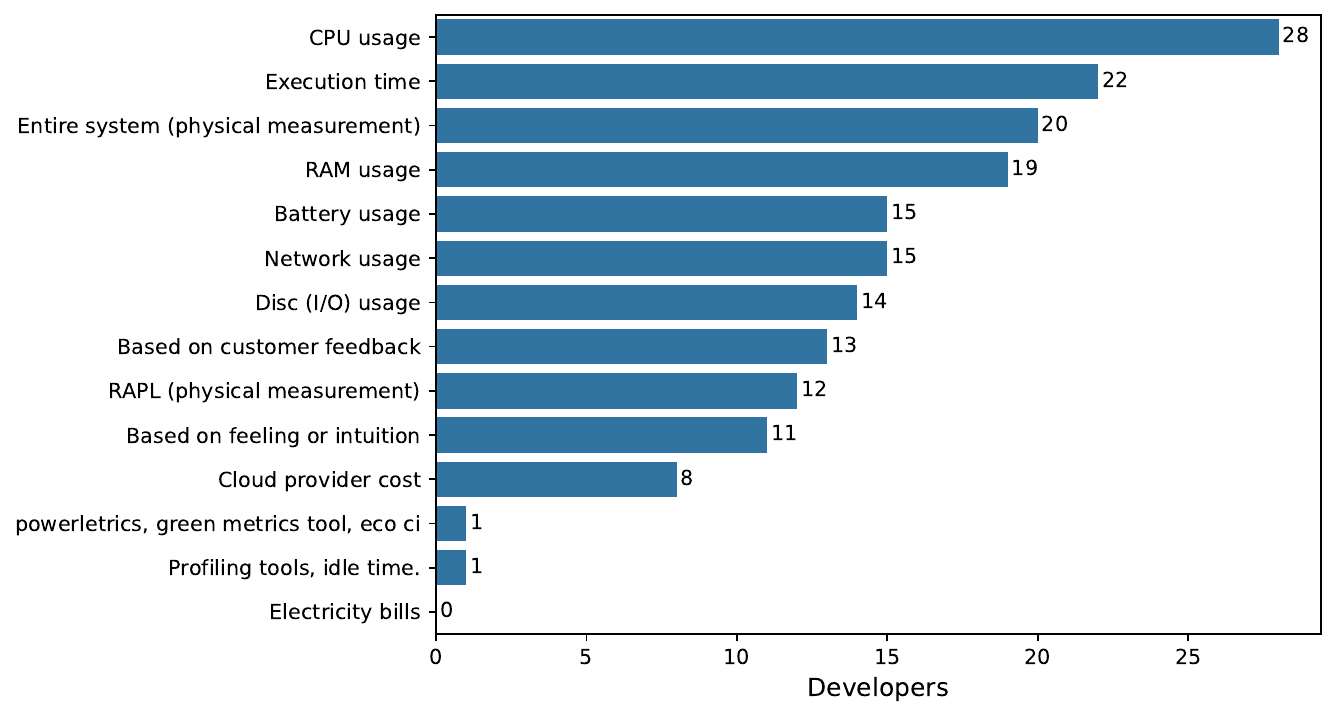}
  \caption{Software energy consumption assessment strategies.}
  \Description{Image showing software energy consumption assessment strategies.}
  \label{fig:energy_assessment_strategies}
\end{figure}

To characterize current practices, specifically, how practitioners estimate or measure the energy consumption of their software, we first focus on the 45 developers who reported that energy consumption already plays some role in their projects. Participants selected known strategies from a curated list derived from literature and could additionally report their own approaches.

Fig.~\ref{fig:energy_assessment_strategies} illustrates that practitioners overwhelmingly rely on \emph{indirect} methods. The most common strategies stated by developers is to approximate energy consumption through CPU utilization (28) and program runtime (22). Direct measurements, such as physical power measurements of the entire system (20) or RAPL-based readings (12), have been stated considerably less often. Additional proxies, including RAM, battery, network, and disk I/O utilization are also used less frequently.
A smaller group of developers relies on subjective or experience-based assessments, including customer \emph{feedback} (13) and \emph{intuition or feeling} (11). These results indicate that where energy is considered, it is often estimated through lightweight or easily observable metrics rather than through exact measurements and instrumentation.

\begin{figure}[t]
  \centering
  \includegraphics[width=\columnwidth]{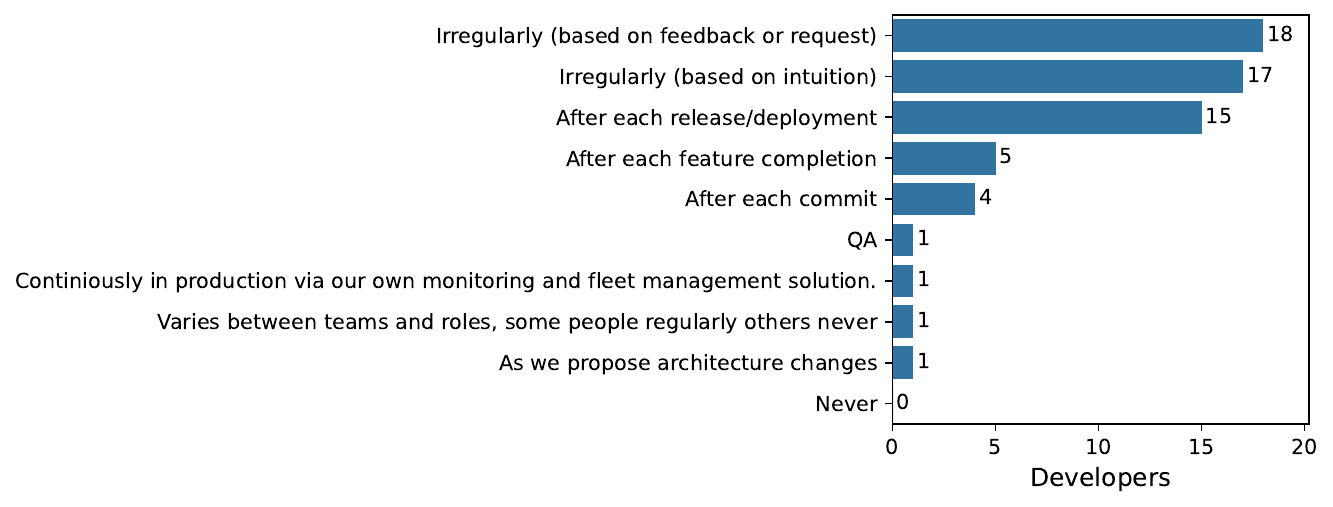}
  \caption{Frequency of assessing software energy consumption in projects.}
  \Description{Image showing frequency of assessing software energy consumption in projects.}
  \label{fig:energy_assessment_frequency}
\end{figure}

Energy consumption is rarely assessed in regular intervals. As shown in Fig.~\ref{fig:energy_assessment_frequency}, the dominant practice is event-driven and irregular: energy consumption is evaluated upon explicit customer request (18) or based on developer intuition (17). More systematic assessments occur only after releases (15) or after feature completion (5). Routine, fine-grained assessments, such as after each commit, are uncommon (4). Only very few participants mention continuous monitoring or QA-driven checks.

\begin{figure*}[t!]
    \centering
    \begin{subfigure}[t]{0.63\textwidth}
        \centering
        \includegraphics[width=\columnwidth]{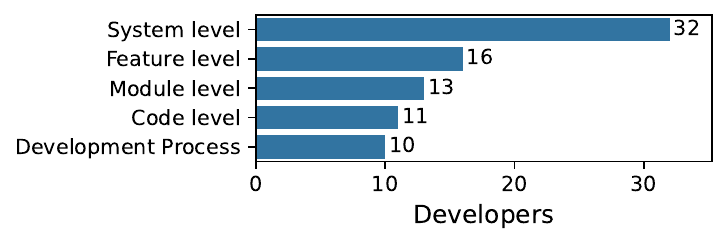}
        \caption{Level of granularity of software energy consumption assessment.}
        \label{fig:granularity_level_assessment}
    \end{subfigure}%
    \hfill%
    \begin{subfigure}[t]{0.34\textwidth}
        \centering
        \includegraphics[width=\columnwidth]{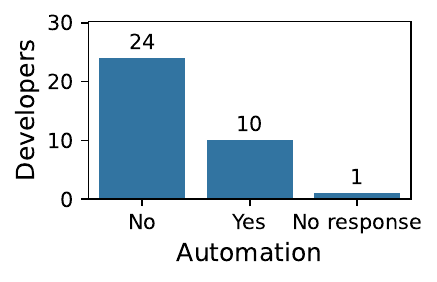}
        \caption{Automation of energy consumption assessment.}
        \label{fig:assessment_automated_or_not}
    \end{subfigure}
    \Description{Image showing automation and frequency of assessing software energy consumption in projects.}
    \label{fig:granularity_plus_automation}
\end{figure*}

Level of granularity of energy consumption assessment. 
Fig.~\ref{fig:granularity_level_assessment} illustrates the level of abstraction at which developers assess energy consumption, ranging from code-level analyses of fine-grained program elements such as individual statements, functions, or classes, up to system-level assessments that capture the energy consumption of the overall system. In addition, development-process–level assessments address the energy consumption related to engineering activities and infrastructure such as CI/CD pipelines or test servers. The results show that developers predominantly examine energy consumption at coarse levels of abstraction, with the system level being most common (32), followed by feature-level (16), module-level (13), and code-level (11) assessments, while assessments tied directly to the development process are least common (10). This preference for higher-level granularity is consistent with the increased complexity and tooling requirements associated with more fine-grained energy measurements.

Despite these limitations and barriers, 10 out of 35 participants reported that they already automate energy assessment in their pipelines, as shown in Fig.~\ref{fig:assessment_automated_or_not}. This illustrates that automated monitoring is achievable in practice and suggests that the main obstacles are not only technical feasibility but rather awareness, education, and perceived relevance.

\subsubsection{Discussion}


Our findings reveal a current practice where energy consumption is acknowledged but not systematically integrated into development workflows. 

Despite our observations from RQ1 that developers frequently justify the absence of energy assessment by citing insufficient tooling, high implementation effort, or conceptual difficulty (Fig.~\ref{fig:no_energy_assessment_arguments}), 10 participants already incorporate automated energy measurements directly into their CI/CD pipelines. The coexistence of these two findings suggests that feasibility may not be the primary limiting factor; instead, the bottleneck may lie in awareness of, and practical knowledge about, available tools and integration approaches, reinforcing the need for improved outreach, education, and demonstrators of successful adoption.

Prior practitioner surveys consistently identify the same tooling and adoption gap across more than a decade~\cite{pang2015programmers, manotas2016empirical, pinto2017energy, hoffmann2025sustainable}: the large majority of developers do not measure energy consumption directly and express a desire for better tooling that remains unmet. Our study corroborates this picture, direct measurement tools such as RAPL\footnote{From the Intel Haswell microarchitecture (2013) onwards, RAPL readings from the CPU package domain can be considered actual hardware measurements rather than model-based estimates, as Intel introduced Fully Integrated Voltage Regulators (FIVRs) directly onto the processor die, enabling on-chip current sensing per power domain~\cite{hackenberg2015energy, burton2014fivr,desrochers2016validation}. It is important to note that this applies specifically to the CPU package domain; the DRAM domain remains partially model-based even on Haswell client systems, and other hardware components such as GPUs, storage, and network interfaces are not covered by RAPL. Sandy Bridge and Ivy Bridge processors (2011–2012) rely entirely on model-based estimation.} (12) remain far less used than CPU utilization (28) and runtime (22) as proxies. Although the research community has produced a growing set of measurement tools and frameworks~\cite{weber2025famlem, aggarwal2015greenadvisor, fieni2024powerapi}, these have clearly not reached mainstream adoption. The practitioner experience of tooling as inaccessible, unfamiliar, or poorly integrated into existing workflows appears structurally unchanged since 2013. The one concrete positive signal is that 10 out of 35 participants already automate energy assessment in CI/CD pipelines, demonstrating that systematic integration is feasible, the barrier is awareness and organizational prioritization, not technical barriers.

Several responses explicitly link energy consumption to performance considerations. Practitioners often assume that improving performance inherently optimizes energy \simplequote{directly proportional to efficiency}{p31} or that performance takes priority \simplequote{faster performance trumped power usage}{p137}. Such statements underscore the importance of proxy metrics for estimating energy consumption. In practice, software energy consumption does not necessarily correlate with proxy metrics, though, including runtime performance~\cite{weber2023twins,chetsa2014exploiting,hao2013estimating}, CPU utilization~\cite{pinto2014understanding,liu2015data}, and disc utilization~\cite{chetsa2014exploiting,kang2016workload}, particularly in heterogeneous or distributed systems.



Our findings illustrate a maturing but still uneven landscape. Energy assessment is technically feasible across a broad range of application domains and deployment contexts, as demonstrated by an active ecosystem of tooling that increasingly addresses even complex virtualized environments. For instance, the Kubernetes-based Efficient Power Level Exporter (Kepler) combines RAPL, ACPI, and NVML readings with an process attribution mechanism to estimate energy consumption at the process, container, and pod levels~\cite{amaral2023kepler}. Recent work has further proposed container-level observability approaches achieving energy attribution within statistical error margins~\cite{pijnacker2025container}, federated learning methods for workload energy prediction across distributed clusters~\cite{saad2025towards}, and systematic characterizations of RAPL-based measurement in virtualized contexts~\cite{raffin2024dissecting}. These developments demonstrate that the technical barrier to energy assessment, while non-trivial, is progressively being lowered.

Yet the landscape remains uneven. The feasibility of energy assessment depends critically on the application domain and the required level of measurement granularity: bare-metal profiling is well-supported, whereas attribution at the container or pod level in shared, multi-tenant environments remains more demanding and is not yet standardized in production practice. Structural incentives, education gaps, and misconceptions further hinder widespread adoption. Integrating energy considerations into software engineering workflows therefore requires not only improved tooling suited to the deployment context, but also organizational alignment and clearer guidance on effective assessment strategies.


\subsection{\RQ{3}: What strategies and development practices are used to mitigate software energy consumption?}\label{eval:rq3}


\subsubsection{Quantitative insights: Mitigation strategies for software energy consumption}

To identify how energy-related issues are addressed by developers once identified, we focus on the responses of 45 developers working in projects where energy consumption already plays an active role. Specifically, we investigated which parts of the software system are affected by energy optimizations and which concrete actions developers have taken when encountering an energy regression.

\begin{figure}[t]
  \centering
  \includegraphics[width=\columnwidth]{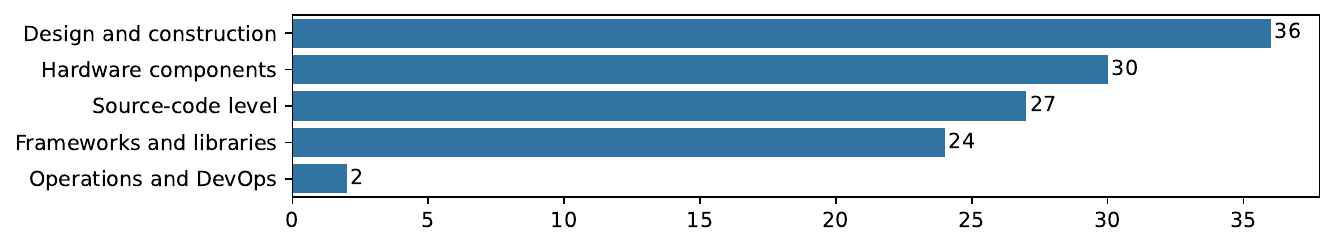}
  \caption{Parts of software system affected by energy-related decisions.}
    \Description{Image showing parts of software system affected by energy-related decisions.}
  \label{fig:influenced_decisions}
\end{figure}

Fig.~\ref{fig:influenced_decisions} shows the parts of a software system that were influenced by energy-related decisions (multiple selections possible). Energy optimizations most frequently affect design and construction decisions (36), indicating that developers predominantly address energy efficiency at architectural and conceptual levels, such as choosing an efficient programming language. Hardware components (30) and source-code level (27) are also commonly involved, reflecting a strong coupling between energy concerns, low-level implementation choices, and execution environments. Frameworks and libraries (24) are slightly less often considered for energy optimizations. This indicates that developers focus more on their own development stack (code and hardware) and may not be aware of energy effects coming from third party products. Surprisingly, operations and DevOps are rarely mentioned (2), indicating that energy-aware practices are weakly integrated into deployment, monitoring, or operational workflows.


\begin{figure}[t]
  \centering
  \includegraphics[width=\columnwidth]{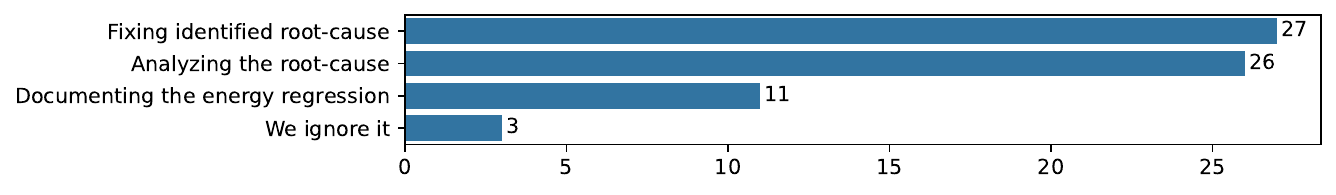}
  \caption{Measures to take when detecting energy regressions.}
    \Description{Image showing measures that are taken when detecting energy regressions.}
  \label{fig:assessment_measures}
\end{figure}

Fig.~\ref{fig:assessment_measures} summarizes the actions developers take after detecting an energy regression. Most participants reported that they analyze (26) and fix (27) the identified root cause, showing that energy regressions are commonly treated as actionable issues rather than being deferred. However, only 11 developers stated that they document energy regressions, suggesting that energy-related issues are rarely institutionalized through systematic reporting or knowledge preservation. A small subset of participants (3) indicated that they ignore energy regressions, highlighting that energy inefficiencies do not always lead to a response, even in projects where energy consumption is considered relevant.

\subsubsection{Open questions: Differences between debugging energy and performance problems}

We further asked participants whether they perceive differences between debugging performance bugs and debugging energy bugs. The responses reveal a notable disagreement. A small majority (17) stated that there is no difference between debugging energy bugs and other non-functional issues such as runtime performance bugs. By contrast, a nearly equal number of developers (15) emphasized that there are differences. 
Participants who reported similarities argued that performance and energy issues are often tightly coupled. One participant stated that \simplequote{performance and energy problems often require similar steps to debug: what changed?; why does it take so much time/energy to do 'x'?}{p115}. Another participant noted that \simplequote{in the GPU stack, power consumption and performance are so closely linked, power regressions are often the same issue as performance regressions}{p128}.
Participants who perceived differences highlighted several distinguishing aspects. First, the identification of energy regressions was described as more challenging than that of performance regressions, as \simplequote{additional steps are required to hopefully find the [energy] problem}{p115}. A concrete example mentioned was that for energy assessment \simplequote{the difference is a change in tools}{p118}. Furthermore, participants emphasized that energy debugging often requires whole-system reasoning, noting that \simplequote{energy debugging tends to be a whole-system problem}{p126}, and that \simplequote{functional bugs are tested on abstraction, energy bugs only exist on actual hardware}{p112}.

Overall, our results indicate that energy-related actions are primarily taken during development rather than being systematically integrated into production or operational phases. While most developers attempt to analyze and fix energy regressions, only 27 out of 45 participants actively engage in remediation, documentation, or sustained energy-aware practices. This raises the question of how the remaining developers respond to energy regressions in practice and whether energy efficiency is treated as a secondary concern when competing with other development priorities.

\subsubsection{Discussion}

Our results reveal a fragmented integration of energy-aware practices into contemporary software development processes. This indicates that energy development practices are much more sensitive to the use case, context, and hardware as performance development practices. That is, many developers assimilate energy regressions into established performance debugging workflows, relying on familiar techniques, tools, and mental models. This approach reflects the technical interdependence of performance, resource usage, and energy consumption and enables developers to reuse existing expertise without introducing new processes.
In contrast, a substantial group of participants identifies energy debugging different to debugging runtime performance problems. These differences arise primarily from limited observability, tool fragmentation, and hardware dependence. Energy consumption cannot be reliably inferred from abstracted execution environments alone and often requires measurements on real hardware, shifting debugging activities beyond traditional software-only boundaries. As a result, energy debugging becomes a demanding, cross-cutting, system-level task that spans software, hardware, and runtime configurations.

The low prevalence of documentation further indicates that energy regressions are rarely embedded into formal development artifacts such as issue trackers or regression reports. This lack of institutionalization limits organizational learning and hinders the long-term integration of energy-aware practices into software engineering processes.

Taken together, our findings suggest that, while energy and performance bugs partially overlap, energy debugging introduces additional methodological and organizational challenges that are not yet fully supported by current development workflows. Addressing this gap requires better-integrated tooling, clearer development practices, and stronger alignment between energy efficiency goals and existing software engineering processes.


\subsection{\RQ{4}: What mental models do practitioners use to relate software energy consumption to other metrics such as runtime performance?}\label{eval:rq4}

Across this study, we repeatedly observed that developers expect software energy consumption to correlate with runtime performance. This assumption surfaced consistently in both quantitative responses and qualitative explanations, indicating a stable mental model in which faster execution is equated with lower energy use. While prior work has examined the relationship between energy consumption and performance empirically, our study is the first to systematically elicit developers’ judgments about the suitability of those proxy metrics and to empirically examine whether these judgments translate into correct expectations about the energy–performance relationship using concrete usage scenarios.

Specifically, we investigate how practitioners conceptualize software energy consumption and how they relate it to other, more accessible metrics in their daily work. As shown in Fig.~\ref{fig:overview_of_question_topics}, we include all participants in this part of the study, independent of whether energy consumption plays an explicit role in their current projects. This decision is grounded in the observation that both groups routinely work with metrics that influence energy consumption: one group of developers (45) does so explicitly with energy optimization in mind, while the other group of developers (89) may change software energy consumption as a byproduct.

\subsubsection{Quantitative insights: Suitability of proxy metrics for software energy consumption.}


\begin{figure}[t]
  \centering
  \includegraphics[width=\columnwidth]{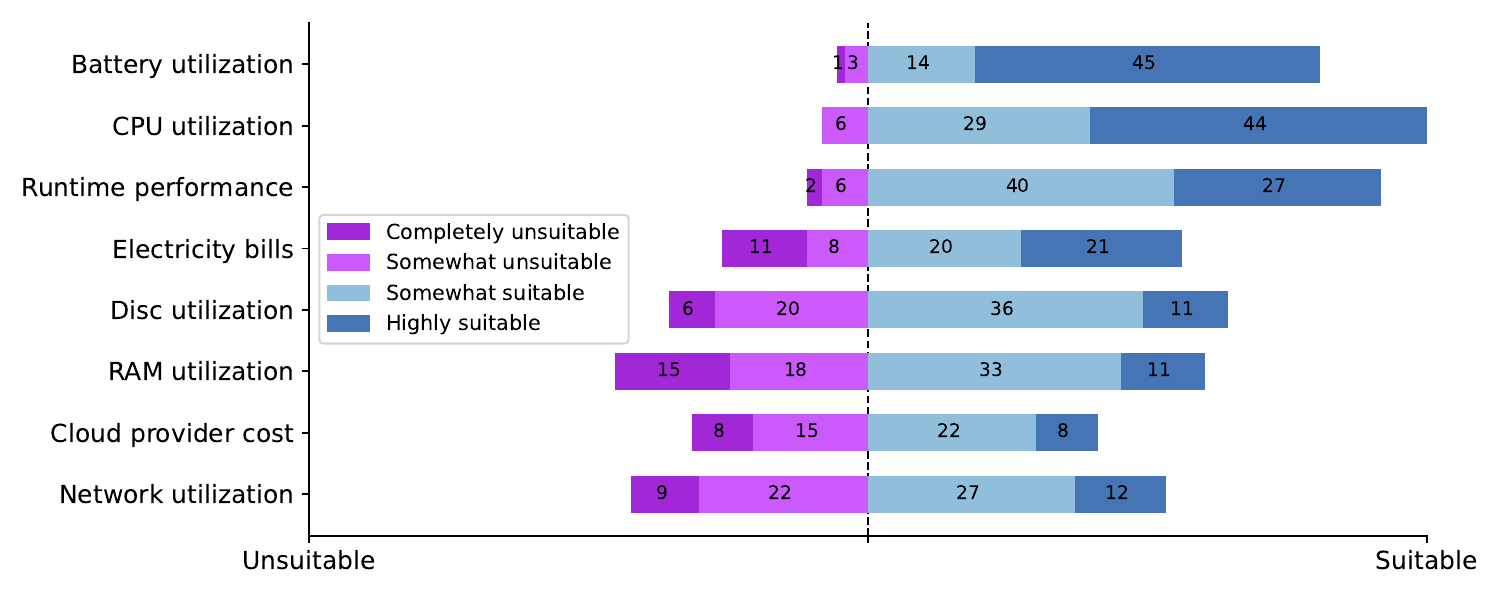}
  \caption{Suitability of proxy metrics for energy consumption.}
    \Description{Image showing suitability of proxy metrics for energy consumption.}
  \label{fig:proxy_metric_suitability}
\end{figure}

Fig.~\ref{fig:proxy_metric_suitability} summarizes the participants’ assessments of the suitability of different proxy metrics for approximating software energy consumption. The selected metrics were compiled from prior studies investigating correlations between energy consumption and system-level or performance-related indicators. 
Three metrics clearly stand out: \emph{battery utilization}, \emph{CPU utilization}, and \emph{runtime performance}. For each of these, a large majority of participants rated them as highly suitable proxy measures for energy consumption. This indicates a mental model in which energy consumption is perceived as tightly coupled to execution time and CPU activity. The same relationship has also been reported in the literature~\cite{li2016automated,babakol2020calm,capra2012measuring}.
Beyond these metrics, the perceived suitability decreases gradually. \emph{Electricity bills} are rated more favorably than \emph{disk utilization} and \emph{RAM utilization}, which is notable given that electricity bills represent an indirect, aggregated outcome rather than a technical system metric. This preference suggests that some practitioners conceptualize energy consumption through economic abstractions rather than resource-level mechanisms.
\emph{Cloud provider cost} and \emph{network utilization} are assessed as the least suitable proxy metrics. Nevertheless, across all proposed metrics, more than half of the participants consider each one at least somewhat suitable for approximating energy consumption.
Overall, these results indicate that developers predominantly reason about energy consumption through performance- and CPU-centric perspectives. This reflects a mental model in which energy consumption is not treated as a first-class property grounded in physical measurement, but rather as an indirect outcome emerging from the optimization of other system-level properties.


\begin{figure}[t]
    \centering
    \includegraphics[width=.8\columnwidth]{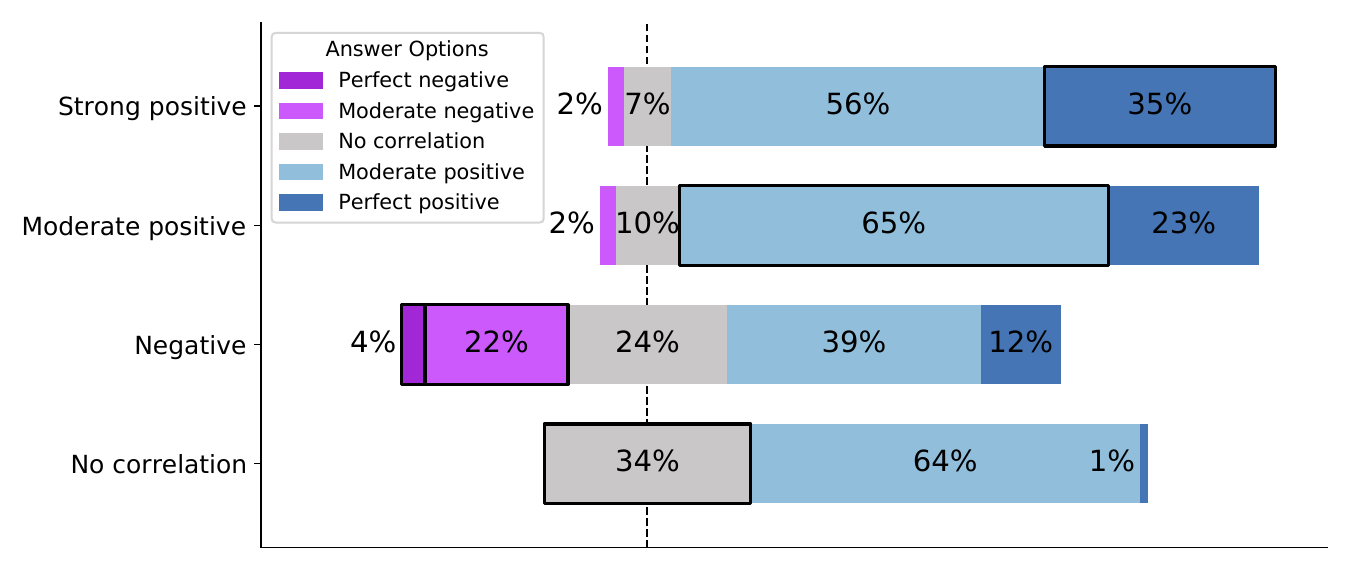}
    \caption{Developers’ expectations of the energy–runtime correlation per scenario (y-axis). The y-axis scenario label indicates the correct answer (also highlighted as black boxes); stacked bars show the distribution of responses.}
    \Description{Image showing expectation of correlation between energy consumption and runtime performance.}
    \label{fig:scenario_distribution}
\end{figure}


To validate whether these proxy-based mental models translate into correct reasoning about energy–-performance dependencies, participants were presented with four real-world scenarios derived from prior empirical studies~\cite{yazdanbakhsh2016axbench,weber2023twins,rucci2015energy}. As discussed in Section~\ref{sec:scenarios}, each scenario varies system conditions and their impact on execution time, and participants were asked to judge the corresponding change in energy consumption. Fig.~\ref{fig:scenario_distribution} summarizes the response distributions.
Across scenarios, participants generally expected energy consumption to increase with longer execution times. This tendency is particularly pronounced in the \emph{strong positive} scenario and the \emph{moderate positive} scenario, where energy consumption and execution time of a video encoding task positively correlate. In both cases, the majority of responses align with the correct classification, reinforcing the dominance of a performance-driven energy model.
The \emph{negative correlation} scenario poses the greatest challenge. Most participants fail to identify the inverse relationship and instead favor a moderate positive correlation. This result highlights a systematic blind spot in the mental models of practitioners. Situations in which performance improvements lead to higher energy consumption are rarely anticipated.
In the \emph{no correlation} scenario responses are more widely distributed. While a substantial fraction of participants correctly identify the absence of correlation, many still expect a positive relationship. This indicates difficulty in recognizing situations where performance changes do not translate into corresponding energy effects, even when such decoupling has been empirically demonstrated.
Taken together, the scenario results confirm that developers predominantly rely on a monotonic performance–energy assumption. While this assumption holds in many common cases, it breaks down in more complex or counterintuitive execution contexts.

\subsubsection{Discussion}


Using proxy metrics to estimate software energy consumption emerges as a widely accepted practice among developers in our study. Performance-related measures such as runtime and CPU utilization are generally regarded as suitable proxies and provide a reasonable starting point for reasoning about energy consumption in many cases. However, related work shows that this assumed positive linear dependency does not hold universally: in specific scenarios (e.g., \emph{negative correlation} and \emph{no correlation}), the relationship between performance and energy consumption breaks down. Identifying and correctly interpreting such cases requires substantial expertise, experience, and domain knowledge. When these assumptions fail, direct energy measurements become necessary to obtain reliable results and to support optimization beyond what proxy-based reasoning can justify.

\begin{figure}[t]
  \centering
  \includegraphics[width=.8\columnwidth]{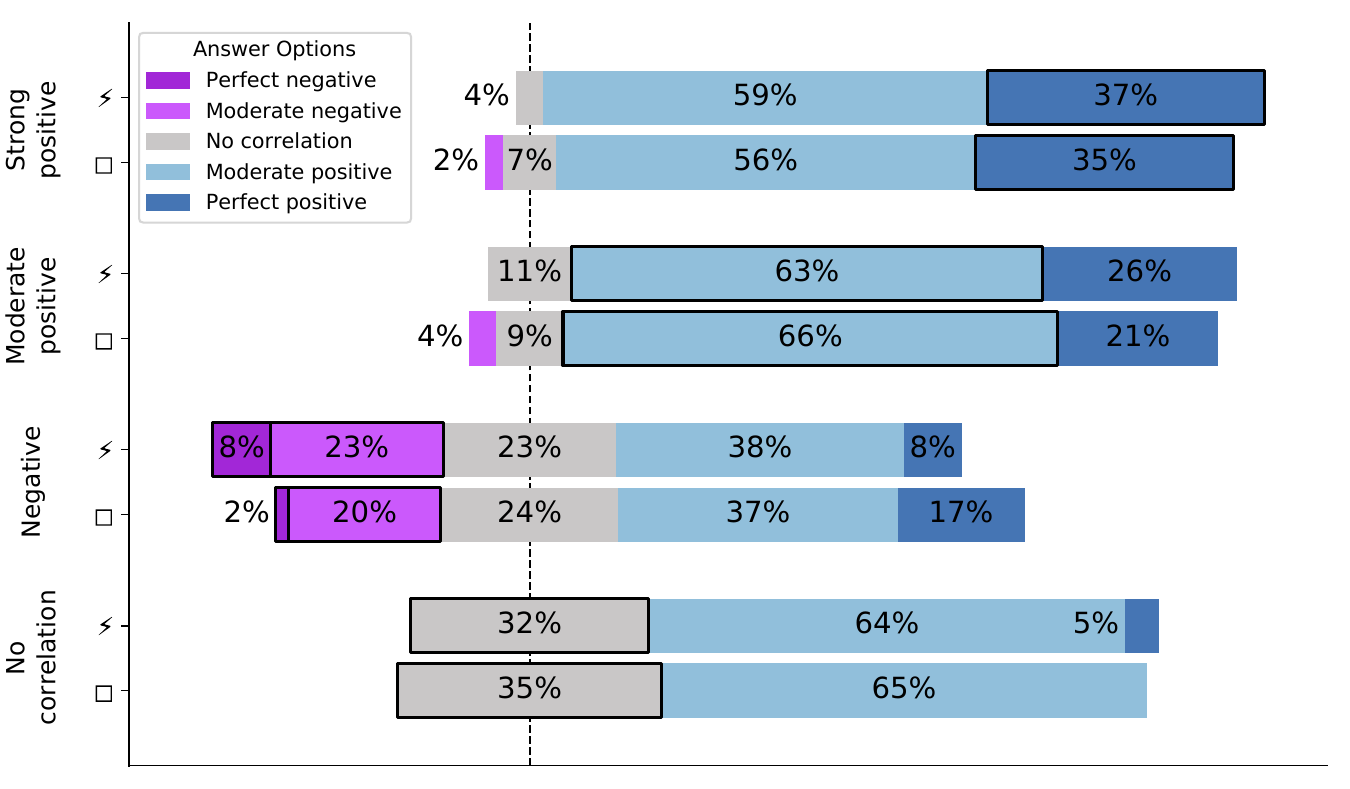}
  \caption{Developers’ expectations of the energy–runtime correlation per scenario (y-axis). The y-axis scenario label indicates the correct answer (also highlighted as black boxes); stacked bars show the distribution of responses. Markers denote developer groups: {\small\Lightning} indicates developers whose current projects treat energy consumption as an explicit concern, and $\square$ indicates developers who do not actively optimize for energy in their projects.}
    \Description{Image showing developer's expectations of correlation between energy consumption and runtime performance.}
  \label{fig:scenario_distribution_relevance}
\end{figure}

Figure~\ref{fig:scenario_distribution_relevance} contrasts developers in energy-relevant projects and those who are not with respect to judging performance--energy relationships across our scenarios. Although developers in energy-relevant projects select the correct correlation slightly more often in some scenarios, the differences are small. A formal significance analysis of expertise and further developer characteristics is presented in Section~\ref{eval:rq4:demographics}.

\subsubsection{Influence of Developer Characteristics on Judgment Accuracy}
\label{eval:rq4:demographics}

The scenario results presented above suggest that developers predominantly rely on a monotonic performance--energy mental model, with systematic difficulties in counterintuitive cases. To investigate whether this pattern is uniform across the developer population or whether certain characteristics predict more accurate judgments, we conducted an exploratory significance analysis. Specifically, we examined four variables that were collected as part of the demographic survey: \textit{energy-aware development expertise} (whether a developer's current project explicitly addresses energy consumption), \textit{years of energy-aware development experience}, \textit{seniority level}, and \textit{application domain}.

\begin{table}[ht]
\centering
\caption{%
    $p$-values from significance tests on ternary accuracy across
    scenarios. Significant results ($p < 0.05$) are highlighted in
    \textbf{bold}. MW~U\,=\,Mann-Whitney U; KW\,=\,Kruskal-Wallis H;
    Sp.\,$\rho$\,=\,Spearman's rank correlation.
}

\label{tab:significance_ternary}
\begin{tabular}{lcccc}
\toprule
\textbf{Characteristic} &
  \textbf{Strong pos.} &
  \textbf{Moderate pos.} &
  \textbf{Negative} &
  \textbf{No corr.} \\
\midrule
Expertise (MW U)                & $0.664$ & $0.689$ & $0.414$ & $0.822$ \\
Years of exp.\ (Sp.\ $\rho$)   & $0.505$ & $\mathbf{0.030}$ & $0.167$ & $0.819$ \\
Seniority (KW H)                & $\mathbf{0.047}$ & $\mathbf{0.012}$ & $0.366$ & $0.199$ \\
\quad Non-Lead vs.\ Lead (MW U) & $\mathbf{0.014}$ & $\mathbf{0.007}$ & --- & --- \\
Domain (KW H)                   & $0.396$ & $0.136$ & $0.114$ & $0.442$ \\
\bottomrule
\end{tabular}%
\end{table}

Table~\ref{tab:significance_ternary} summarizes the results. Overall, the analysis yields little evidence that the collected developer characteristics systematically differ between groups. Neither energy-aware expertise nor application domain produce significant results in the overall aggregation or across any individual scenario. Notably, years of energy-aware development experience yields a significant result in S2 ($\rho = 0.260$, $p = 0.030$), though not in the remaining scenarios or in the overall aggregation. This indicates that the difficulty of judging energy--performance correlations is broadly shared across developer profiles, consistent with the pattern already observed in Fig.~\ref{fig:scenario_distribution_relevance}.

The only characteristic that provides hints toward a systematic effect is \textit{seniority}. In the two positive-correlation scenarios (S1 and S2), the Kruskal-Wallis test reaches significance ($p = 0.047$ and $p = 0.012$, respectively), with small-to-moderate effect sizes ($\eta^2 = 0.068$ and $\eta^2 = 0.113$). Dunn's all-pairs post-hoc test, however, found no significant pairwise differences between Junior, Senior, and Lead groups under Bonferroni correction. A targeted follow-up contrast, collapsing Junior and Senior developers into a single \textit{Non-Lead} group and comparing against Lead developers, yielded significant results for both scenarios ($p = 0.014$ and $p = 0.007$, rank-biserial $r = 0.338$ and $r = -0.308$). In scenario S2, Lead developers show a notably lower median accuracy score than Non-Lead developers, suggesting better recognition of the moderate positive correlation. This contrast was, however, identified post-hoc after observing that pairwise tests were non-significant, which is an important caveat for statistical validity.

For \textit{years of energy-aware development experience}, a significant Spearman correlation emerges in S2 ($\rho = 0.260$, $p = 0.030$), suggesting a weak positive association between experience and accuracy in the moderate positive correlation scenario. No such association is found in the remaining scenarios or in the overall aggregation. This isolated finding is likely not independent of the seniority effect observed in S2: years of experience and seniority level tend to co-vary, and the two signals may therefore reflect the same underlying factor rather than distinct predictors of judgment accuracy.

To investigate whether any combination of characteristics jointly predicts judgment accuracy, we additionally conducted an ordinal logistic regression for each scenario. The regression results do not yield findings beyond what the per-characteristic analyses already indicate: no predictor reaches significance, and the model explains only a small fraction of the variance in accuracy (McFadden $R^2 = 0.063$). The full regression output, including coefficients, standard errors, and odds ratios, is published in our supplementary material~\footnote{Supplementary Web page: \supplMaterial.}.

Taken together, our findings point to a broader knowledge gap that affects developers regardless of whether they work in energy-relevant projects. While proxy metrics serve as foundational cognitive tools for reasoning about software energy consumption and explain their widespread and persistent use in practice, their failure modes reveal unexploited energy-saving potential. This highlights the need for energy-aware tool support that can contextually qualify proxy-based reasoning by automatically detecting scenarios in which commonly used proxies such as performance are likely to become unreliable, particularly in parallel, memory-intensive, or communication-heavy workloads, and by explicitly signaling when direct energy measurements are required to enable effective optimization. From an educational perspective, systematically exposing developers to counterexamples that violate common performance-energy assumptions may further support the development of more robust and situation-aware mental models.

\subsection{\RQ{5}: What strategies, resources, or incentives could support practitioners in better managing software energy use?}


\subsubsection{Open question: Challenges in adopting energy-aware software development}

\begin{figure}[t]
  \centering
  \includegraphics[width=\columnwidth]{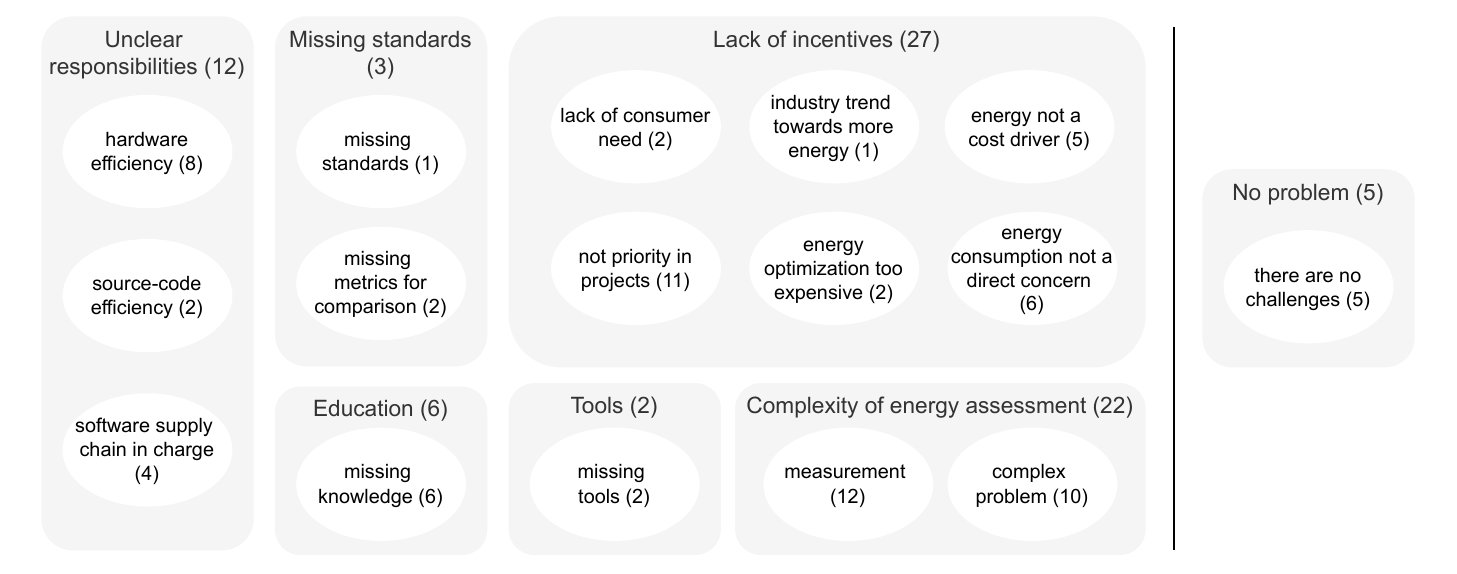}
  \caption{Developer's statements on challenges that hinder energy-aware software development practices. Survey question: ``Reflecting on your previous answers, what challenges do you face with energy consumption in your software projects?''}
  \Description{Image showing developer's statements on challenges that hinder energy-aware software development practices.}
  \label{fig:adoption_challenges}
\end{figure}

The participants provided open-ended responses describing obstacles that hinder the adoption of energy-aware software development practices in their projects and, in some cases, reflected on the absence of such obstacles. Across all responses, developers articulated 72 statements describing challenges, alongside 5 responses explicitly denying the existence of challenges. We clustered these statements into thematic categories using card sorting (see Fig.~\ref{fig:adoption_challenges}). We discuss now each category, starting with the most frequently mentioned challenges.

\textit{Lack of incentives (27).}
The most dominant cluster concerns missing incentives for engaging in energy-aware software development. Developers consistently described situations in which energy consumption does not influence decision-making, as it is perceived as economically irrelevant or subordinate to other priorities. Several participants emphasized that energy is not considered a cost driver, stating that \simplequote{energy costs don't really matter}{p41}. Others described organizational constraints that lower prioritization of energy optimization, for example, noting that \simplequote{it is not easy to even get approval for any optimization tasks}{p73}. A contributing factor appears to be a low market demand, as developers observed that energy efficiency \simplequote{would only interest a fraction of the intrinsically motivated customer base}{p19}. Together, these statements indicate that, in the absence of economic or contractual incentives, energy-aware practices remain difficult to justify within existing project structures.

\textit{Unclear responsibilities (12).}
This category captures uncertainty about who is responsible for addressing software energy consumption. Several developers shifted responsibility toward hardware choices, arguing that \simplequote{total energy consumption [is] usually predetermined by the choice of hardware}{p36}. Others pointed to the software supply chain, particularly \simplequote{inefficient third party libraries}{p33}, as a major source of inefficiency beyond their direct control. In contrast, a smaller subset emphasized developer responsibility at the source-code level, identifying inefficiencies such as \simplequote{the main culprit is simply doing unnecessary or redundantly repeated processing}{p36}. These perspectives reveal fragmented responsibility attribution, which complicates systematic adoption of energy-aware practices.

\textit{Educational problems (6).}
A subset of participants highlighted deficits in knowledge and expertise related to software energy consumption. These responses describe organizational contexts in which teams lack foundational understanding, for example, \simplequote{in big companies, there are lots of teams with no knowledge of energy consumption}{p81}. Others emphasized methodological uncertainty, stating that \simplequote{root cause analysis [is] a bit of a guessing game to some level}{p118}. Such statements suggest that even when motivation exists, insufficient conceptual and practical knowledge limits effective action.

\textit{Missing tools (2).}
Closely related, but less frequently mentioned, are explicitly technical barriers. These responses focus on missing or inadequate tooling and infrastructure, with participants stating that \simplequote{when power concerns need to be addressed, training, tooling, and documentation is inadequate}{p112} and highlighting that there is \simplequote{missing tooling and metrics to identify energy inefficient parts in the software}{p36}. Although less prevalent, these statements point to concrete limitations within current development environments. Interestingly, this is where most research is focused on, indicating a mismatch of research and problems in practice.

\textit{Missing standards (3).}
Some developers identified the absence of standards as a barrier to energy-aware development. This includes both procedural and measurement-related aspects, such as \simplequote{there is a lack of standardization (in the sense of ISO standards) as a foundation}{p18}, and calling for a \simplequote{quickly evaluable and traceable metric}{p15}. Without shared standards, energy-related efforts remain difficult to benchmark, compare, and communicate.

\textit{Energy assessment (22).}
A large cluster of responses highlights the complexity of assessing software energy consumption. Developers described difficulties in identifying optimization targets, noting that \simplequote{specific areas to optimize is hard}{p61}. Others emphasized the challenge of causal attribution, as \simplequote{linking cause and effect in power regressions can be difficult, as there are multiple asynchronous units all involved in a workload. Their interactions can be very complex, even for relatively simple differences in input}{p128}. Measurement itself was repeatedly described as problematic, with statements such as \simplequote{not always clear how to reduce energy consumption, or how to measure it}{p79} and \simplequote{it's just hard to measure, multi threading in particular already makes performance hard to predict in many cases, but in terms of energy consumption its even worse to predict because there's a completely new dimension of efficiencies to factor in}{p199}. These responses underscore that energy assessment constitutes a technical and methodological challenge, since identifying and attributing root causes of energy behavior is inherently difficult.

\textit{No problem (5).}
Finally, a small group of participants reported no perceived challenges, responding with statements such as \simplequote{there aren't any challenges}{p222} or \simplequote{None?}{p32}. Given their brevity and lack of elaboration, these responses may reflect disengagement or fatigue rather than a well-grounded assessment, and they stand in contrast to the detailed challenges articulated by the majority of participants.

\subsubsection{Open Question: Addressing identified challenges}

\begin{figure}[t]
  \centering
  \includegraphics[width=\columnwidth]{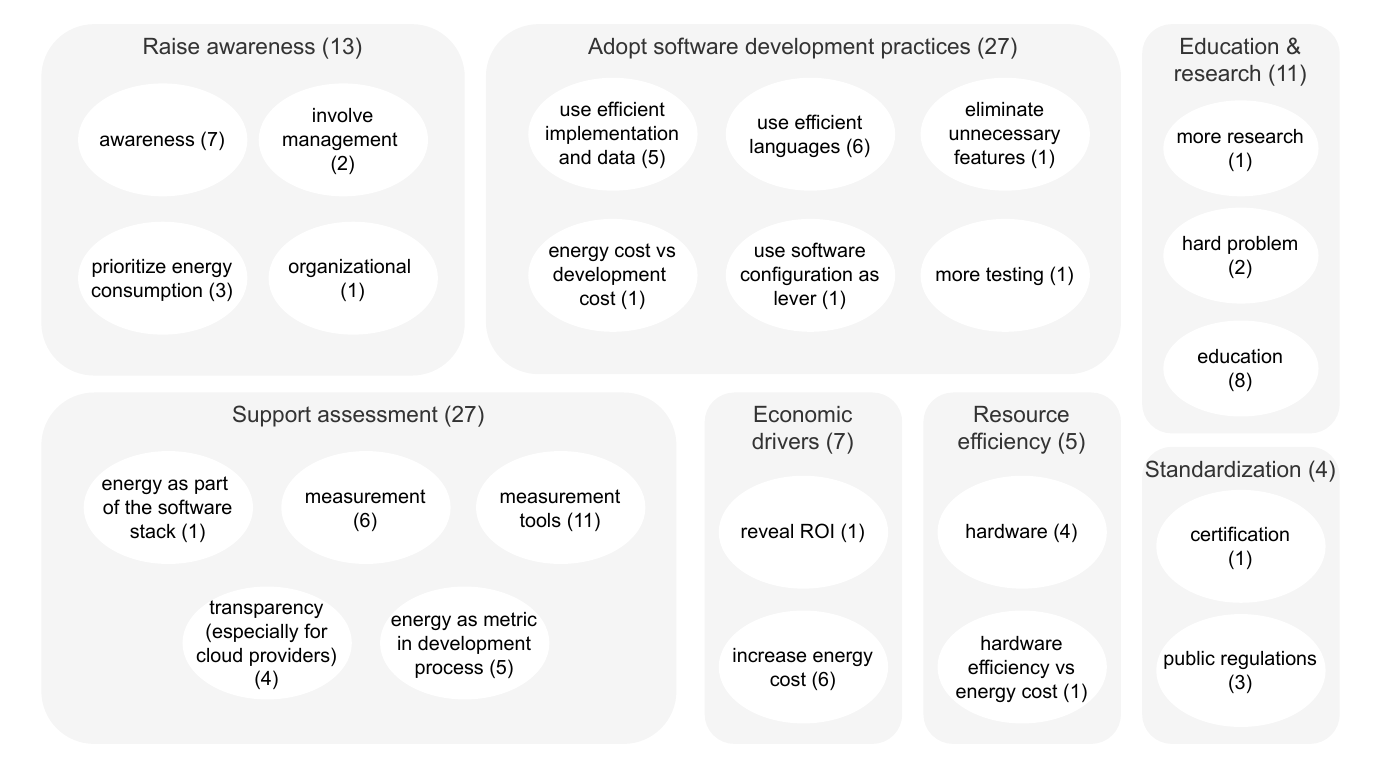}
  \caption{Developer's statements on solutions to overcome challenges that hinder energy-aware software development practices. Survey question: ``What do you think needs to be done to address the challenges of energy consumption?''}
  \Description{Image showing developer's statements on solutions to overcome challenges that hinder energy-aware software development practices.}
  \label{fig:adoption_solutions}
\end{figure}

To understand how the identified challenges might be addressed, we analyzed the developers’ responses to the follow-up question on potential solutions. In total, participants articulated 94 solution-oriented statements, which we clustered into categories shown in Fig.~\ref{fig:adoption_solutions}. We discuss how these proposed solutions relate to the previously identified challenges and which challenges remain unaddressed.

\textit{Economic drivers (7).}
Given the prominence of missing incentives, several developers proposed economic mechanisms as a central lever. Suggestions included making benefits explicit by \simplequote{demonstrating economic value and clearly highlighting the effectiveness of one’s own work in improving software energy efficiency}{p43}, as well as increasing cost pressure, such as \simplequote{there must be monetary motivation to write efficiently, otherwise it won't happen}{p67}. One participant argued that energy costs are currently negligible in certain contexts, stating that \simplequote{within the operational costs for datacenters, the actual energy usage is relatively low priority, human labor costs and hardware failures are among the largest expenses}{p222}. This perception contrasts with established cost structures in modern data centers, as energy consumption and cooling alone account for the majority of cost in modern data centers~\cite{poess2008energy,rong2016optimizing,aryee2025enhancing}. This suggests that misinformation or outdated assumptions may contribute to the lack of incentives identified in the results.

\textit{Adoption of software development practices (27).}
Many solution-oriented statements address unclear responsibilities by embedding energy awareness into established development practices. Participants suggested choices at the language and implementation level, such as \simplequote{hardware needs to be used more efficiently. Especially higher level languages can improve a lot in that regard.}{p55} or \simplequote{use natively compiled languages.}{p114}. Others focused on coding practices, emphasizing \simplequote{write faster, simpler code which doesn't draw as much power}{p132}. These statements reflect a common mental model that closely links performance optimization with energy efficiency. However, changing the programming language alone is insufficient, as inefficient or wasteful implementations can equally be realized in languages that are generally considered energy efficient. Consequently, these responses point toward the need to embed energy-aware practices across the software development life cycle rather than relying on isolated technical choices. This interpretation aligns with prior work calling for the systematic integration of green coding techniques throughout the software life cycle, as also proposed by Hoffmann et al.~\cite{hoffmann2025sustainable}.

\textit{Education \& research (11).}
Educational and research-oriented solutions directly address the knowledge gaps identified in the challenges. Developers called for additional research, noting that \simplequote{research [is needed]; the topic remains extremely vague so far}{p15}. Others emphasized insufficient understanding among practitioners, stating that \simplequote{many developers have too shallow of an understanding [of software energy consumption]}{p55}. The perceived difficulty of the topic was also highlighted, as effective optimization requires \simplequote{a deep understanding of how the target machine works (CPU/GPU architecture) and know how to profile a program with various utilities}{p100}. These responses suggest that targeted education and research are important, rather than short-term fixes.

\textit{Support for energy assessment (27).}
Many statements of the developers target the complexity of energy assessment. Participants advocated for making energy a first-class concern in the software stack, for example through \simplequote{energy consumption as a metric provided directly by the operating system}{p181}. Measurement was repeatedly emphasized, with statements such as \simplequote{measurement is the most important thing; it needs to be easy to measure energy consumption}{p101} and calls for \simplequote{more accurate and granular measurements}{p128}. Developers further requested better tooling, including \simplequote{open-source measurement tools}{p17}, \simplequote{better profilers or APIs to access energy values}{p28}, and \simplequote{widely available tooling for estimating energy consumption}{p79}. Especially in cloud environments, participants emphasized the need for greater visibility into energy usage, calling for \simplequote{transparency of the energy consumption of cloud services}{p18}. Finally, participants proposed to improve integration, suggesting to \simplequote{provide and integrate tooling and metrics throughout all development stages}{p36} and to assess energy impacts early, \simplequote{for example in a manner comparable to test code coverage}{p7}. Overall, these proposals suggest that making energy assessment feasible in practice necessitates coordinated advances in education, tooling, and research, alongside the technical integration of energy metrics into development workflows.

\textit{Resource efficiency (5).}
Some developers viewed energy efficiency primarily as a hardware issue. Statements emphasized the role of infrastructure with demands, such as \simplequote{more focus on power efficient CPUs, it's a hardware issue}{p131}. Others expressed skepticism about the impact of software-level efforts, arguing that \simplequote{the most radical improvements will come from wide adaptations of better, more energy-efficient hardware}{p70}. These perspectives partially align with the earlier attribution of responsibility to hardware choices and, in doing so, shift responsibility for energy awareness away from software developers.

\textit{Standardization (4).}
Standardization was proposed as a means to address missing benchmarks and comparability. Suggestions included \simplequote{expansion of certification schemes, for example the Blue Angel~\footnote{Blue Angel: \url{https://www.blauer-engel.de/en/productworld/software}} for software.}{p11} and broader regulatory approaches, such as \simplequote{long-term political prioritization, for example through further expansion of renewable energy and public infrastructure.}{p11}. At the same time, concerns were raised about unintended consequences, as one participant cautioned that \simplequote{regulations are likely to aggravate the existing issues with quality, cost, and timely completion of projects even further}{p37}. This reflects ongoing tension between formalization and development agility.

\textit{Raising awareness (13).}
Finally, several responses focused on awareness towards the energy consumption of software systems. Developers stressed the need for general awareness, noting that \simplequote{there needs to be more awareness what it costs to have the software running}{p41}, and for \simplequote{increase consumer awareness}{p178}. Management was identified as a key actor, with calls for \simplequote{a shift in management mindset; not focusing exclusively on producing new features, but also considering code quality}{p212}. Others argued that prioritization must be explicit, stating that \simplequote{I don't believe you can get developers to care unless efficiency is part of the task spec}{p134}. At the same time, several participants framed energy awareness as an individual responsibility of developers, suggesting that progress depends on \simplequote{an attitude by developers to consider power consumption in their products}{p64} and observing that experienced developers are already \simplequote{conscious by heart when it comes to performance and energy consumption}{p66}. Taken together, these statements reveal a tension between self-perceived developer awareness and motivation on the one hand, and organizational, managerial, and market-level barriers on the other, with developers positioning themselves as willing actors constrained by external priorities.

\subsubsection{Discussion: Challenges and solutions in adopting energy-aware software development}

Taken together, the proposed solutions directly address several of the most prominent challenges identified in the results. In particular, assessment complexity is met with strong calls for improved measurement, tooling, and integration into development workflows. Educational deficits are explicitly acknowledged and paired with demands for research and training. Unclear responsibilities are partially addressed through lifecycle-oriented practices and tooling that embed energy concerns into everyday development activities.
However, structural incentive gaps remain only partially resolved. While economic drivers and awareness are frequently proposed, many solutions depend on organizational, management, or market-level changes that extend beyond individual developers’ control. Similarly, standardization and regulation emerge as necessary but contested areas, indicating unresolved trade-offs between comparability, enforcement, and development flexibility. Overall, the results suggest that while technical and methodological challenges are increasingly understood, aligning incentives and responsibilities across stakeholders remains a central open problem for energy-aware software engineering practice.

\subsection{Implications for Practitioners}

Our findings suggest several concrete directions for practitioners seeking to improve energy-aware software development in their projects.

\paragraph{Push for direct measurements where necessary.}
While proxy metrics such as runtime performance and CPU utilization are reasonable starting points for energy estimation in many scenarios, our results demonstrate systematic blind spots in proxy-based reasoning. Specifically, memory- and network-intensive workloads, parallel execution, and configurable systems can exhibit negative or absent energy--performance correlations that proxies fail to capture. In these contexts, direct energy measurements, such as RAPL-based readings or system-level power metering, are necessary to obtain reliable results and to avoid optimization decisions based on incorrect assumptions.

\paragraph{Treat energy assessment as technically feasible, not optional.}
Energy measurement is widely perceived as difficult and time-consuming, yet a notable subset of practitioners in our study already incorporates automated energy assessments into CI/CD pipelines. This demonstrates that the primary barrier is not
technical infeasibility but rather a lack of awareness of available tools and integration approaches. Practitioners should leverage existing measurement frameworks, uch as RAPL, PowerAPI, or FAMLEM~\cite{weber2025famlem}, and treat automated energy checks analogously to test coverage gates: as a standard part of the development workflow rather than an exceptional activity.

\paragraph{Make energy consumption visible in the development process.}
Our results show that energy regressions are rarely documented and that energy assessment is almost entirely absent from operations and DevOps workflows. A practical first step is to treat energy regressions as first-class development artifacts: tracked in issue systems, reported in regression pipelines, and monitored in production alongside existing performance metrics. Making energy effects visible at decision points that developers already act upon, which lowers adoption barriers without requiring entirely new processes.

\paragraph{Anchor performance-as-proxy reasoning in its limits.}
Runtime performance and CPU utilization are already near-universally used as energy proxies, and for CPU-bound, single-threaded workloads this is largely justified. Practitioners should, however, be aware of the conditions under which this mental model breaks down. Teams working with parallel, memory-intensive, or communication-heavy workloads should explicitly validate proxy-based assumptions through targeted energy measurements, rather than assuming that performance improvements translate into corresponding energy reductions.

\paragraph{Build organizational and economic justification.}
Missing management interest and absent customer demand are the most frequently cited barriers in our study. Practitioners advocating for energy-aware practices should frame energy efficiency in terms directly relevant to project stakeholders: reduced cloud and infrastructure costs, extended battery life in mobile and embedded contexts, or compliance with emerging regulatory requirements. Concrete, project-specific return-on-investment arguments are more likely to unlock organizational support than abstract sustainability goals.

\subsection{Threats to Validity}
\label{validity}

\subsubsection{Internal validity}
A primary threat to internal validity arises from our participant recruitment strategy. We disseminated the survey through professional social networks (e.g., LinkedIn) and selected online communities, including technology-specific subreddits. This recruitment approach may introduce selection bias, as participation was voluntary and likely attracted developers with an existing interest in software quality, performance, or energy efficiency. As a result, participants may exhibit higher baseline awareness of energy-related concerns than developers in the broader population.
While these factors limit claims about prevalence, they do not invalidate the qualitative insights reported in this study. Moreover, our study found a large portion of developers who do not engage in energy assessment and optimization, so we were able to collect responses from diverse viewpoints. Our analysis focuses on identifying the current practices of energy-aware software development, mental models energy abstraction, and challenges rather than estimating an general overview. Within this scope, the collected responses provide internally consistent and richly articulated perspectives on energy-aware software engineering coming from experts in the field.

Another possible threat to validity may be the card sorting process. To reduce the possible threat it was conducted collaboratively by two authors and iteratively refined to ensure consistency of identified categories. While alternative categorizations are possible, the identified themes are well-supported by multiple independent statements across participants. Given the exploratory and descriptive nature of the study, our conclusions are appropriately framed as analytical insights rather than causal claims.

\subsubsection{External validity}
The external validity is limited by the characteristics of the sampled population. By recruiting participants primarily from professional networks and specialized developer communities, our study reflects the perspectives of relatively experienced developers operating in environments where performance and efficiency concerns may already be more salient than in general-purpose software development. Consequently, our findings may not directly generalize to less experienced developers, teams working predominantly with higher-level abstractions, or domains where energy consumption is rarely discussed.
However, energy-aware software engineering practices are more likely to emerge, be discussed, or be contested in communities where performance and resource efficiency are already relevant concerns. Studying these contexts allows us to identify both advanced practices and persistent barriers that may also affect broader adoption. Nonetheless, caution is required when extrapolating the results beyond similar professional and technical settings.

A further limitation concerns the composition of recruited communities. Targeting subreddits such as r/cpp and r/rust may over-represent systems programmers with a prior interest in low-level performance and resource efficiency. The sample of participants is therefore diverse relative to a single-company or single-language study but is not a representative cross-section of the software engineering profession at large.

\subsubsection{Construct validity}
A threat to construct validity arises from the scenario-based questions used in RQ\textsubscript{4}. Three of the four scenarios are grounded in empirical findings reported in the authors' own prior work~\cite{weber2023twins}. 
However, the underlying energy-performance effects described in the scenarios are not unique to that work and have been independently reported across multiple studies~\cite{yazdanbakhsh2016axbench, rucci2015energy, chetsa2014exploiting, dzhagaryan2014impact, li2016evaluating}, lending broader empirical support to the scenario premises.

A related concern is that the answer options for the scenarios might linguistically pre-load the correct response. In particular, the option ``consumes double energy'' implies a precise linear relationship, which is a stronger claim than the \emph{strong positive correlation} scenario strictly warrants. Participants who reason carefully may reject this option not because they misunderstand the energy-performance relationship, but because they correctly recognize that an exact doubling is unlikely in practice. This could lead to an underestimation of correct responses in that scenario. Future work shall consider formulating answer options that separate directional judgments from magnitude claims more clearly.

\section{Conclusion}
\label{conclusion}

Our study provides a comprehensive empirical view of how software energy consumption is perceived, assessed, and acted upon by professional developers in contemporary software engineering practice. By means of a mixed-methods survey of 134 practitioners, we reveal a nuanced landscape in which energy efficiency is widely acknowledged as relevant but inconsistently accounted for.

A central finding is the distinction between explicit and implicit engagement with software energy consumption. Only a minority of developers work in projects where energy efficiency is treated as a first-class requirement. Nevertheless, the majority routinely optimize performance-related properties, such as runtime performance, CPU utilization, or memory footprint, that directly affect energy consumption. This observation challenges the common narrative that developers largely ignore energy concerns, instead highlighting that energy-related effects are often addressed indirectly and unintentionally.

Furthermore, our results show that developers overwhelmingly rely on proxy metrics to reason about energy consumption. Runtime performance and CPU utilization dominate both assessment practices and mental models. While this proxy--based reasoning is frequently valid, our scenario-based evaluation demonstrates systematic misconceptions: developers tend to overgeneralize a positive correlation between performance and energy consumption and struggle to anticipate negative or absent correlations. Even developers with prior experience in energy-aware projects exhibit persistent biases in ambiguous scenarios. These findings suggest that proxy metrics, while indispensable in practice, can foster overconfidence and incorrect expectations when contextual limitations are not made explicit.

Organizational and structural factors strongly shape whether energy-aware practices are adopted. Developers report missing incentives, unclear responsibility, and limited customer or management interest as dominant barriers. At the same time, a notable subset of participants already integrates automated energy assessment into CI/CD pipelines, demonstrating that technical feasibility is not the primary obstacle. Instead, gaps in awareness, education, and dissemination of existing solutions appear to play a decisive role.

Taken together, our findings point to several implications for research and practice. 
First, energy-aware software engineering should explicitly acknowledge and support the widespread use of proxy metrics, while clearly communicating when and why these proxies fail. 
Second, tooling should focus on seamless integration into existing workflows, making energy effects visible at decision points developers already act upon. 
Third, education and training need to address common misconceptions about energy–performance relationships and emphasize counterexamples that violate intuitive assumptions. 

By grounding these insights in empirical evidence from practitioners, this paper contributes to a realistic understanding of the current state of energy-aware software development and outlines concrete directions for closing the gap between awareness and systematic practice.



\bibliographystyle{ACM-Reference-Format}
\bibliography{bibliography}

\end{document}